\documentstyle[11pt,aaspp4]{article} 
\newcommand{\kms}{km~s$^{-1}$}
\newcommand{\kmsM} {km~s$^{-1}$~Mpc$^{-1}$}
\newcommand{\subsun}{\mbox{$_{\odot}$}}
\newcommand{\etal}{{\it et al.\/}}

\lefthead{Cohen {\it et al.} }
\righthead{Redshift Survey}

\slugcomment{submitted to The Astrophysical Journal}

\begin{document}

\title{Caltech Faint Galaxy Redshift Survey VIII:
	Analysis for the Field J0053+1234\altaffilmark{1}}

\author{Judith G. Cohen\altaffilmark{2},
        Roger Blandford\altaffilmark{3},
        David W. Hogg\altaffilmark{3,4,5}
        Michael A. Pahre\altaffilmark{2,5,6} \&
        Patrick L. Shopbell\altaffilmark{2}}

\altaffiltext{1}{Based in large part on observations obtained at the
	W.M. Keck Observatory, which is operated jointly by the California 
	Institute of Technology and the University of California}
\altaffiltext{2}{Palomar Observatory, Mail Stop 105-24,
	California Institute of Technology, Pasadena, CA \, 91125}
\altaffiltext{3}{Theoretical Astrophysics, California Institute of Technology,
	Mail Stop 130-33, Pasadena, CA \, 91125}
\altaffiltext{4}{Current Address:  Institute for Advanced Study, Olden Lane, Princeton, NJ \, 08540}
\altaffiltext{5}{Hubble Fellow}
\altaffiltext{6}{Current Address:  Harvard-Smithsonian Center for Astrophysics, 
	60 Garden St., Mail Stop 20, Cambridge, MA \, 02138}

\begin{abstract}
The results of a spectroscopic investigation of a complete sample of
objects with $K_s<20$~mag in a 2 by 7.3~arcmin field at J005325+1234 are
reported.  
Redshifts were successfully obtained for 163 of the 195 objects in the sample; these redshifts
lie in the range [0.173, 1.44] and have a median of 0.58 (excluding 24 Galactic stars).
The redshift identifications are believed to be almost complete for $z < 0.8$.
Approximately half of the galaxies lie in five narrow redshift features 
with local velocity dispersions of ${\sim}300$~\kms.
These narrow redshift ``peaks'' are primarily populated both by absorption--line galaxies 
and the most luminous galaxies in the sample, 
although the incidence of emission lines
in the luminous galaxies increases with redshift.
The estimated dynamical masses of these redshift peaks, and the 
sky distribution of the
galaxies within them, appear similar to groups or poor clusters of galaxies
in the local universe at various stages of virialization.
Some groups of galaxies therefore form at epochs $z > 1.5$,
and the galaxies in such groups appear to be coeval and to
show little sign of ongoing star formation.
The galaxies outside the redshift peaks are also clustered, albeit more weakly, 
are less luminous and more frequently exhibit strong emission lines.
These ``isolated'' galaxies therefore appear, on average, to form 
stars at later epochs than the strongly clustered galaxies.
The galaxy SEDs derived from our $UBVRIK$ photometry are also very closely
correlated with the galaxy spectral types and luminosities.  These results
have strong implications for the analysis of redshift surveys at intermediate
redshift.

The sample is used to investigate the evolution of the combined 
galaxy luminosity function back to $z=0.8$.
No significant change is found in the characteristic 
luminosity $L^{\ast}$ and only 
weak color changes are detected, consistent with passive evolution. The blue
galaxy
luminosity function is more dwarf-rich than the red galaxy 
luminosity function.  No
significant change in the comoving density is found in this sample
out to $z \sim 1.4$ assuming that the objects without redshifts
(16\% of the sample)
are galaxies essentially all of which have $z > 0.8$.  This 
suggests that mergers are not important among the objects in this sample.

A population of extremely red objects with $(R-K)>5$~mag exists in the infrared-selected sample;
all four such objects with redshifts are found to be absorption line galaxies with $z \sim 1$.
Most of the very red objects therefore appear to be  galaxies
with $z \gtrsim 1$ which are not heavily reddened by dust.

A measure of the UV extinction at 2400\AA\ for the emission line
galaxies of a factor of two is obtained, implying only modest
UV extinction in high redshift star-forming galaxies.
\end{abstract}


\section{Introduction \label{introduction} }

We observe that galaxies evolve.  The number of very faint (and, presumably, distant) galaxies on the
sky is at least $8\times 10^{10}$ (\cite{williams96}), 
roughly thirty times more than would be naively predicted on the basis of the comoving
volume of the Universe and the local bright-galaxy luminosity function
(\cite{loveday92}; \cite{marzke94}; \cite{lin96a}; \cite{gardner97};
\cite{hogg97}; \cite{ratcliffe98}).  
In addition, faint galaxies are bluer (\cite{koo92}; \cite{smail95}),
smaller (\cite{smail95}; \cite{griffiths94b}), and more irregular 
(Griffiths \etal\ 1994a,b;\markcite{griffiths94a,griffiths94b}
\cite{glazebrook95}; \cite{driver95}; \cite{abraham96}; \cite{odewahn96}) 
than bright galaxies at the present day.

In order to translate these observations into a 
description of the physical evolution of the whole galaxy
population with cosmic time, it is necessary to 
understand the galaxy redshift distribution.
For this reason much telescope time has been
devoted to large redshift surveys which allow us to 
measure the luminosities and ages of statistical samples
of galaxies.  In addition, they also explore, directly, 
the variation of the star formation rate, the chemical abundance,
and non-thermal activity with cosmic time, which are strong markers of the evolutionary history
of galaxies. Furthermore, these surveys can be used,
globally, to study the clustering and spatial distribution
of faint galaxies.

The Caltech Faint Galaxy Redshift Survey, (henceforth CFGRS),
is designed to measure the properties of field galaxies in the redshift
interval $0.3\lesssim z\lesssim1.3$.  The survey is described
in Cohen et al (1998\markcite{cohen_apjsupl98}), where it is contrasted
with the many other recent redshift surveys of field galaxies.
The CFGRS uses complete samples to a fixed limiting magnitude in a particular 
bandpass within a small solid angle on the sky.
Spectra are obtained for every object in the sample with the Low
Resolution Imaging Spectrograph (henceforth LRIS) (Oke \etal\ 1995) on
the 10~m Keck Telescope.  

This paper presents spectroscopic results from one field located at J005325+1234, 
the central region of which is part of the Medium Deep Survey (\cite{griffiths94a})
and among the deepest fields imaged with HST prior to the Hubble Deep Field.
The field measures $2 \times 7.3$~arcmin$^2$ with a statistical 
sample containing 195 
infrared-selected objects complete to $K = 20$~mag, of which 
24 are spectroscopically confirmed Galactic stars 
and 32 cannot be assigned spectroscopic redshifts. (21 of these have spectra). 
A preliminary report on the field J0053+1234 was given by Cohen \etal\ 
(1996a\markcite{cohen96a}) where it was shown that roughly half of the galaxies
are located in dense groups.
The sample definition and $UBVRIK$ photometric catalog are presented in a companion 
paper (\cite{pahre98apjs}), while the redshifts are given in a
second companion paper, Cohen \etal\ (1998\markcite{cohen_apjsupl98}).
The properties of the Galactic stars are discussed in Reid \etal\ (1997\markcite{reid97}).
Large scale structure will be discussed, in the context of this survey, in a 
forthcoming paper (\cite{cohen99b}).  
A redshift survey of another region, the Hubble Deep Field (\cite{williams96}, 
henceforth the HDF), will be the subject of the next paper in this
series (\cite{cohen99a}).

This paper is structured as follows.
The galaxy luminosities and colors in the observed frame
are discussed the \S\ref{observed}. 
The extremely red galaxies in our sample are discussed in \S\ref{veryred}.
Then the galaxy spectral energy distributions
(SEDs) are constructed in the rest frame and it is 
shown that these are strongly correlated with the
galaxy spectral classes in \S\ref{luminosities}.
On this basis, redshifts are provisionally assigned for the 32 galaxies 
in our complete sample whose spectra lack adequate line features. 
After using our SEDs to deduce the ultraviolet extinction for 
strong emission line galaxies with $z \sim 1.2$ in \S\ref{uvext},
we then discuss the overall
distribution of galaxies in redshift and comoving volume.    
Next, we distinguish the clustering of the different spectral classes 
in \S\ref{spectra_groups}
and demonstrate that the prominent groups contain apparently older galaxies.
The limited morphological information that we have on our sample is presented
along with some predictions in \S\ref{morphology}.
Finally, our conclusions are collected and discussed in 
\S\ref{summary} along with some suggestions for future research.

We adopt the values $H_0 = 60$~\kmsM and 
$\Omega_M = 0.3$ with $\Lambda = 0$ throughout.
The present age of the universe in this cosmology is 13~Gyr and the current
cosmological density is $2\times10^{-30}$~g cm$^{-3}$.

\section{Galaxy Properties in the Observed Frame\label{observed} }

\subsection{Hubble Diagrams \label{hubblediagrams} }

\placefigure{fig1}

The conventional way to represent the results of redshift surveys is with 
a Hubble diagram.  Three Hubble diagrams, $B(z)$, $R(z)$ and $K(z)$  
are presented as Figures~1a,b,c for the redshift sample. 
The four different galaxy spectral classes defined in
Cohen \etal\ (1998\markcite{cohen_apjsupl98})
(``$\cal{E}$'' for emission line dominated galaxies,
``$\cal{A}$'' for galaxies showing only absorption features,
``$\cal{C}$'' for galaxies with composite spectra
and ``$\cal{Q}$'' for AGNs)
are distinguished on these diagrams. (A few galaxies were not detected at
$B$ and these are ignored in Figure~1a.) We also exhibit the magnitudes 
of the 32 objects 
from the galaxy sample that do not have redshifts when these are detected.  
As is usual, we ignore reddening internal to the galaxies themselves 
and from the IGM. 

As the rest wavelengths are blueshifted with respect to our six 
bandpasses, it is 
also conventional to correct the magnitudes for this effect using k-corrections.
These have been tabulated by Poggianti (1997\markcite{poggianti97}) who conveniently
separates this purely spectral effect from the change due to the evolution
of the stars that we observe locally.  Together, these two effects are 
called ``passive'' evolution.
(We have corrected for our different filters and cosmology.)
The spectral evolution predicted by Poggianti for
a galaxy with $L = 2L^{\ast}$ with no evolution is 
also exhibited on the Hubble diagrams for local galaxies with
SEDs of an elliptical, Sa, and Sc type.\footnote{The SED for the local elliptical is
from Bruzual \& Charlot (1993\markcite{bruzual96}).}  The
passive evolution predictions from Poggianti (1997\markcite{poggianti97})
are also shown in Figure~1a,b,c, again for
a $2L^{\ast}$ galaxy.   
\footnote{The k-correction assigned to an elliptical galaxy as computed from
the Worthey (1994\markcite{worthey94}) and  
Bruzual \& Charlot (1996\markcite{bruzual96}) models (as provided in \cite{leitherer96}) 
show good agreement with each other, and with the Bruzual \& 
Charlot (1993\markcite{bruzual93}) model described
above.
They also show good agreement with the ``empirical'' approach to estimating
$k$--corrections by Cowie \etal\ (1994\markcite{cowie94}), which was based
on broad-band photometry of nearby elliptical galaxies.
These $k$--corrections differ substantially by $\sim 0.4$~mag from the 
Poggianti (1997\markcite{poggianti97}) models at $z \sim 0.6$ for the $K$ filter, 
suggesting that the latter models are problematical at near--infrared 
wavelengths.}

An enormous range in galaxy luminosities is apparent in Figure~1. 
For intermediate redshift galaxies observed in the $R$ band, there 
is a five magnitude spread in 
luminosities. The second notable feature of this diagram is the concentration 
of roughly half of the galaxies, predominantly ``${\cal A}$'' galaxies, 
in redshift clumps.  A third peculiarity is the dearth of galaxies, especially
``${\cal A}$'' galaxies, for $z\lesssim0.3$ and $0.8\lesssim z\lesssim1.1$. Fourthly,
we note that Poggianti's (1997) passive evolution tracks
do not appear to represent the observed run of
galaxies, particularly at $B$ and $R$.   Including the
evolutionary correction term for passive evolution leads to luminosities
that decrease as $z$ increases, while
use of just
the k-correction term (i.e. no evolution) is in fact better.

It has long been argued that elliptical galaxies in clusters of galaxies
show only passive evolution, at least out to $z \sim 0.5$ 
(\cite{aragon93}; \cite{kelson97}; \cite{pahre98thesis}; \cite{postman98}).
Evolutionary brightening of disk galaxies in the field 
is also quite modest as measured
using surface photometry (\cite{schade96}; \cite{barger98a}) or the
Tully-Fisher relation (Vogt \etal\ 1996\markcite{vogt96}, 1997\markcite{vogt97}),
reaching only $0.5$~mag in the $B$--band at $z \sim 0.5$.  More recently
Hogg (1998) in an analysis of the HDF has shown that there evolution
of $L^\ast_B$ is at most modest out to $z\sim1$.  
Hamilton (1985\markcite{hamilton85})
has demonstrated that the amplitude of the 4000\AA| jump is 
approximately constant for bright field elliptical galaxies out
to $z \sim 0.8$.

\subsection{Galaxy Colors\label{galcolors}}

A conventional view of galaxy colors is given in 
Figures~2a,b,c, which 
show the dereddened galaxy colors $U-R$, $R-K$ 
and $U-K$ for the full sample as a function of
redshift.  Galaxies without a detection at $U$ are not plotted.
The same symbols are used to indicate the galaxy spectral
types as appear in figure~1.
The thin and thick lines represent the predictions of the Poggianti (1997\markcite{poggianti97})
models for no-evolution and for passive evolution respectively. 
For galaxy colors, only differences of the values of the
evolutionary corrections computed in each of two colors for
the passive evolution models of Poggianti (1997) are relevant,
while in comparing galaxy luminosities with predictions from models, the
values themselves are at issue.  Thus the effects of any
errors in the evolutionary corrections in Figures~2a,b,c
are more subtle than those
seen in the Hubble diagrams (Figure~1a, b, c).

Figure~2 shows overall agreement with the color predictions from
no-evolution models, as has been noted earlier by, for example,
Oke, Gunn, \& Hoessel (1996\markcite{oke96}) for galaxies in clusters with $z \sim 0.5$ and
in the CFRS (\cite{crampton95}), but still many minor concerns persist.  
In each panel of this figure, the range of the observed galaxy colors
(equivalent to the range of galaxy SEDs, i.e., the range of
star formation histories considered valid for galaxies)
is somewhat larger at all redshifts than the range predicted by the models.

The evolutionary corrections are better studied by a more carefully 
defined sample of galaxies in clusters where the galaxy spectral type 
can be more tightly constrained.  
Such an approach has been taken by Pahre (1998\markcite{pahre98thesis}).

\placefigure{fig2}

\subsection{The Very Red Objects \label{veryred} }

At high galactic latitude there is a population of faint, very red objects.
Among the main sample (\cite{pahre98apjs}) there are 19 objects 
with $(R-K) \ge 5$~mag---three of which have $(R-K) > 6$~mag---producing a surface density on
the sky of $\sim$1.3 arcmin$^{-2}$.
Ignoring Galactic stars, this class of objects comprises 11\% of the number 
counts for $K<20$~mag.
  
The spectroscopically confirmed stars in the total sample all have $(R-K)<4.6$~mag,
and are mostly M dwarfs.
While the reddest galactic M dwarfs reach $(V-K) \approx 6$~mag (\cite{leggett92}), 
such extremely red stars are not common in magnitude limited samples.  

Lawrence \etal\ (1995\markcite{lawrence95}), among others, advocate that a significant fraction of galaxies at high
redshift are dusty.  However we do not find significant UV extinction among the 
emission line galaxies in our sample (\S\ref{uvext}), and hence do not believe that the many of the very
red objects in our sample without measured redshifts are heavily reddened blue galaxies.

Four of the 19 very red objects have measured redshifts.  
All four of them are classified as ``$\cal A$'', with $0.78 < z < 1.23$; 
three of the four have $z > 1$.
Persson \etal\ (1993\markcite{persson93})
have speculated that such objects are passively-evolved 
elliptical galaxies with $z > 1$, while 
Graham \& Dey (1996\markcite{graham96})
suggest that they are reddened star forming galaxies.
%
%
%
The very red objects for which we were successful in measuring a redshift
support the hypothesis that such objects
are high redshift elliptical galaxies, specifically 
that they are galaxies with $z \gtrsim 1$ that are not heavily reddened by dust.

\section{Spectral Energy Distributions\label{luminosities} }
\subsection{Computation of the Rest Frame Spectral Energy Distributions}

For each galaxy in our redshift sample, we can construct a rest frame
spectral energy distribution (SED). We do this in a slightly non-standard way.
For each galaxy, we compute the luminosity per $\ln\nu$,
$L\equiv\nu L_\nu$, as a function of rest frequency using the six photometric magnitudes
of Pahre \etal\ (1998\markcite{pahre98apjs}).
The calibrations for absolute flux were adopted from 
Bessell (1979\markcite{bessell79}) for
$U$ through $R$, and were calculated from the material 
in Pahre \etal\ (1998\markcite{pahre98apjs}) for $I$ and for $K_s$.
A selection of raw SEDs is displayed
in Figure~3.  In constructing raw SEDs, we 
interpolate linearly between measurements.  The nominal error on each measured point is
dominated by systematic effects. We estimate this to be $0.2$ in magnitude
or $0.08$ in $\log L$.  For a variety of reasons discussed in   
Pahre \etal\ (1998\markcite{pahre98apjs}), many photometric 
measurements have much 
larger errors and these are shown as vertical bars. In addition, several photometric
measurements are only $2\sigma$ upper limits and these are also designated in Figure~3.

\placefigure{fig3}

We next transform these raw SEDs into corrected SEDs by interpolation and extrapolation
through the inaccurate measurements and upper limits and extension redward to 2.2~$\mu$
using the observed $I-K$ colors.  (We believe this procedure to be quite robust
because the near infrared spectra of most galaxies are well represented as power
laws.)  We also introduce two new spectral bands, $P$ and $Q$ 
at rest frequencies 
$\log\nu=15, 15.1$ in the 
vacuum ultraviolet which are directly observed in most cases. 
(We extrapolate
to these frequencies in the lowest redshift galaxies.) This operation leaves us with
continuum spectra and a set of 8 rest spectral luminosities $L_\alpha$, where $\alpha=$
$K, I, R, V, B, U, P, Q$. (The biggest concern about this procedure is the large gap
in $\log\nu$ between $K$ and $I$. This can introduce a systematic steepening of the derived
infrared rest spectra for high redshift galaxies as the observed 
$I$ band corresponds
to rest $B$ where the rest SED can be intrinsically curved.  However, 
none of our results is strongly dependent upon this extrapolation.)

These SEDs are listed in Table~1 and 
are exhibited for a selection of 
galaxies in the redshift sample in Figure~4. 
There are some striking (though not unexpected) regularities.  
The absorption line galaxy SEDs exhibit quite red infrared and 
ultraviolet spectra.  For convenience we define 
two spectral indices in the rest frame. $\alpha_{IR}$ is the spectral index,
$-d\log L_\nu/d\log\nu$, measured between the 
rest $B$ and $K$ bands and $\alpha_{UV}$ is the 
corresponding quantity between the $Q$ and $B$ bands. 
(Again our results are quite robust to
this choice, which maximizes the tightness of the correlations
which follow.) If we restrict our
attention to ``${\cal A}$'' galaxies with high quality
redshifts as defined in \cite{cohen_apjs98} and with $z < 0.8$, 
then nearly all the galaxies are located in the 
region defined by $\alpha_{IR}>1,\alpha_{UV}>3$ and almost all of 
``${\cal E}$'' galaxies with high quality redshifts
lie outside this region (Figure~5).
(The association of a hard ultraviolet continuum with emission lines is not, of course, a surprise.  
However, this correlation does demonstrate that internal reddening
to be discussed in \S\ref{uvext}
is not a big factor in these galaxies.)
We can then use this correlation to define two SED classes for the whole redshift sample,
including the ``${\cal C}$'' spectral class, which cover the whole two-color 
plane, and those of low quality (mostly ``${\cal A}$''
class).  We call these classes ``old'' ($\alpha_{IR}>1,\alpha_{UV}>3$) and ``young'', although
the latter may also be rejuvenated, consisting of an older population plus
a recent starburst.

\placefigure{fig4}

\placefigure{fig5}

The correlation of $\alpha_{IR}$ with $L_K$ is shown in Figure~6 for the
same set of galaxies with high quality redshifts that are displayed
in Figure~5.  A strong correlation between $L_K$ and $\alpha_{IR}$ is
apparent; the most luminous galaxies have redder spectral indices.
There is also a clear separation with galaxy spectral type.  More luminous
galaxies tend to be of ``$\cal{A}$'' spectral class, while the least luminous
ones are those that show strong signs of recent star formation
(the ``$\cal{E}$'' galaxies).

The tight correlation between galaxy SED shapes (defined from 2400\AA\
to 2.2$\mu$ in the rest frame) and
galaxy spectral classes assigned on the basis of the presence
or absence of key diagnostic features ([OII]3727, H+K, [OIII]5007, etc)
is one reason why photometric redshift techniques
such as that of Connolly \etal\ (1995\markcite{connolly95}) works reasonably
well at least out to $z \sim 1$ for high precision photometric data sets.

\subsection{Luminosity -- Volume Diagram \label{bolometric} }

In order to compare galaxies at different redshifts (and also with galaxies studied in other
surveys), it is necessary to define a fiducial luminosity (or absolute magnitude).
We again follow convention and label the galaxies by their 
rest $B$ luminosities.
We do this in a slightly non-standard way,  
eschewing tabulated
k-corrections and models of galaxy evolution. Instead, for each galaxy,
we compute the spectral energy distribution (SED) of the luminosity per $\ln\nu$,
$\nu L_\nu$, as a function of rest frequency and extrapolate to the rest $B$ 
frequency, (nominally $\log\nu_B=14.83$), from the red so as to define a 
rest $B$ luminosity
$L_B \equiv \nu L_\nu(\nu_B)$. We do this because of the presence of variable
4000\AA\ breaks just to the blue of the $B$ band. 

We normalize the galaxy 
luminosity to the local, fiducial luminosity $\log (L_B^\ast)=36.86$ W, or $M_B^\ast
=-20.8$ mag, or $\log(L_B /1 {\rm L}_{\odot B})=10.52$
(\cite{bingelli88}) in our cosmography. 
(Note that 
$\log(L_K^{\ast}=36.91 {\rm W} \equiv M_K^\ast=-24.6$ mag, (\cite{mobasher93}); \cite{cowie96} and so a median $L^\ast$ galaxy
should have $\alpha_{IR}\sim1$, somewhat bluer than our median galaxy, just as might be expected
from a $K$-selected sample.)
We also estimate the total luminosity in the wavelength interval $0.4\mu<\lambda<2\mu$,
which is probably a fair measure of the bolometric luminosity, by assuming 
a power law fit to the SED so that
$L_{{\rm tot}}=T(\alpha_{IR})L_B$, where $T(0),T(1),T(2)=0.9,1.6,3.6$ respectively.  
For most of the galaxies in our sample, $0.5<T(\alpha_{IR})<2.5$.

Figure~7 shows the luminosity for each galaxy 
in our redshift sample as a function of both comoving volume and redshift 
separated by spectral class.  It is immediately apparent 
that the luminosity function does not evolve strongly out to $z\sim0.8$
and that there is a serious deficit of galaxies with 
$0.8\lesssim z\lesssim 1.3$, an issue to which we turn next.

\placefigure{fig10}

\subsection{Galaxies without Redshifts\label{nozgalaxies}}

The sample contains 32 objects believed to be galaxies 
for which we cannot assign 
redshifts.  However, we are fairly confident that these are not
normal galaxies with $z\lesssim0.8$ because 
normal galaxies with strong emission lines in that redshift range are detected
with identifiable line(s) all the way up to the survey limit.  While
faint absorption line galaxies in this regime are more problematic, the
characteristic shape of the 4000\AA\ break is still visible for such galaxies
with $z < 0.8$, at least to $R \sim 24$ mag, corresponding to $(R-K) > 4$ mag
at the survey limit.  Although it is possible that we could
be dealing with a new population of galaxies that exhibit neither 
emission nor absorption lines, similar to BL Lac objects, we regard
this as quite unlikely because this population would have to be spectrally
heterogeneous and to have maximum luminosity at a given redshift
that evolved so as to track our survey limit.  Conversely,
if these galaxies are at high redshift ($z > 2$), they would be anomalously
luminous.  Their median $B$ luminosity at
$z\gtrsim1.3$ would be ${\gtrsim}3L_B^{\ast}$, requiring a discontinuous
jump in the luminosity function.  The most plausible explanation is that most
of these galaxies have $0.8{\lesssim}z{\lesssim}1.3$.

\subsection{Ultraviolet Extinction\label{uvext}}

The amount of absorption in the ultraviolet suffered by 
the Lyman-break  galaxies that are found by Steidel \etal\ 
(1996\markcite{steidel96}) and others
is a matter of great current interest and debate.
It affects the inferred luminosities of these objects as well as parameters
calculated from these luminosities, such as the
rate of star formation and the metallicity in the Universe as a function of
cosmic time deduced by Madau, Pozzetti \& Dickenson (1998\markcite{madau98}).  Meurer \etal\ (1997\markcite{meurer97}) and \
Sawicki \& Yee (1998\markcite{sawicki98})
claim a mean extinction $\ge 10$ at 1500\AA\ for $z \sim 3$ galaxies,
while Trager \etal\ (1997\markcite{trager97}) advocate a much smaller value.  
The results from the first searches at sub-mm 
wavelengths (\cite{blain98}, \cite{barger98b}, \cite{eales98}) have been used
to suggest that there is a substantial population of dust enshrouded
star forming galaxies at high redshift.  At low redshifts ($z < 0.2$) 
the mean UV extinction from dust at mid-UV wavelengths
appears to less than a factor of 2 from analysis of H${\alpha}$ emission
line strengths (Tresse \& Maddox 1998\markcite{tresse98}), while
Heckman \etal\ (1998\markcite{heckman98}) have used IUE spectra to
study the UV continuum of local starburst galaxies to conclude that their
average extinction is a factor of 10 in the UV for
a solar metallicity galaxy, with extinction decreasing in galaxies
that are more metal poor.

The closest analogs to the high-$z$ galaxies in our sample are the
galaxies with strong emission lines.
We attempt to use our SEDs, with their very broad wavelength coverage, 
to constrain the extinction within ``$\cal{E}$ ''galaxies at 
$0.9 < z < 1.5$, and then argue that dust
should build up with time as star formation and metal formation
proceeds, and hence any limit we obtain should apply even more rigorously
at higher $z$. 

We obtain a limit on UV extinction through
comparing the index of the power law fit to the UV SED 
($\alpha_{UV}$) to that of
the UV continuum observed for nearby starburst galaxies and that 
predicted by models of starburst galaxies such as those of
Leitherer \& Heckman (1995\markcite{leitherer95}).   
Since our definition of $\alpha_{UV}$ is the
power law index between 
rest frame $B$ and $Q$ (2380\AA), and there may be substantial
contribution from older stars at $B$, we use the power law index defined
from our rest frame SEDs between $U$ and $Q$
instead.  The median index for the 10 such galaxies with high
quality redshifts in our sample
with strong emission lines in that redshift range is 1.3.  
The power law fit for extreme unreddened starbursts is $-2.1$ 
(\cite{calzetti94}, \cite{leitherer95})
(in $F_{\lambda}$, while
we are using $F_{\nu}$) over
the regime from 1200 to 3000\AA, and corresponds to a very young
burst with O stars dominating the UV flux.  This difference in the power
law index,
if attributed completely to extinction, implies a median
extinction at 2380\AA\ of a factor of 1.8.  
The actual value must be smaller as at least some of 
the $U$ flux is coming from the older stellar population in these galaxies
at $z \sim 1.2$; their infrared colors indicate the presence of a
substantial population of stars older than O stars.  

We transform the extinction at 2380\AA\ to a value at the
region of interest, 1500\AA.  There
are a variety of extinction laws
in use reviewed by Calzetti, Kinney \& 
Storchi-Bergmann (1994\markcite{calzetti94}) and Calzetti 
(1997\markcite{calzetti97}) 
as the wavelength dependence of the extinction
in the UV appears to depend on environment.  These 
are essentially identical at wavelengths redder than
2400\AA, but differ in ``grayness'' at shorter wavelengths.
Thus our median extinction  at 2380\AA\ of a factor of 1.8 converts to a
median extinction at 1500\AA\ of between a factor of 2 and 3, assuming
no change in the dust properties of galaxies between $z \sim 1.3$ and
$z \sim 3$.  If anything, we expect the dust content to be lower
at $z \sim 3$. 

Because the sample of galaxies from which this value was obtained is selected
at $K$, there can be essentially no highly reddened objects with strong emission lines
in this redshift range that are not included in the sample.
Furthermore because the spectroscopy is reasonably complete, especially
for objects with strong emission lines, there can
be no additional heavily reddened starbursts in this
redshift range concealed among the objects in the sample.

In conclusion, unless the dust in high redshift galaxies has unprecedented
properties, we believe 
that searches based on flux in the rest frame UV
of high redshift galaxies
do not suffer substantial selection effects due to large amounts
of UV extinction. We suggest the presence of modest mean
UV extinctions, a factor of $\sim$3 at 1500\AA.  This is in 
good agreement with a preliminary determination
by Pettini \etal\ (1998\markcite{pettini98}) using similar techniques
applied to a small sample of $z \sim 3$ galaxies themselves of 1 -- 2 mag at 1500 \AA.

\subsection{Age estimation\label{ages}}

Estimation of the ages of galaxies in the sample is important for
understanding their evolution, especially since, as will be shown
below, the galaxies thought to be physically associated because they
lie in the same peak in the redshift distribution show evidence for
having similar ages.  In addition, at the highest redshifts in
the sample, age determinations put constraints on the cosmological world
model, or at least on the earliest epochs of galaxy formation.
Recently, such constraints have been derived from a very red $z=1.55$
galaxy selected by its radio emission (Dunlop \etal\ (1996\markcite{dunlop96}); Spinrad \etal\ (1997\markcite{spinrad97}); 
Heap (1998\markcite{heap98}).  However, such objects may be atypical in
important ways since they contain active nuclei which may affect
interstellar media, stellar initial mass functions (IMFs), and
SEDs.  The redshift sample presented
here contains a significant number of $z>1$ sources, not because
magnitude selection is efficient for this, but merely because the
sample is fairly complete.  For this reason, we also expect this $z>1$
galaxy sample to be typical, or at least more typical than a
radio-selected sample.

The standard method for age-dating a (fairly) young galaxy involves
assuming that the near ultraviolet spectral energy distribution is
not severely affected by extinction (see \S\ref{uvext}) and is
dominated by stars near the main-sequence turnoff of the most recent
significant burst of star formation.  
Because there is the possibility of earlier bursts, the age so
derived is only a lower limit.

This age-dating technique can, in principle, be performed by comparing
the near-ultraviolet spectral index $\alpha_{UV}$ (as defined above)
of the galaxies in the sample to those of hot stars.  
Conversion of observed blue colors into a rest-frame near-ultraviolet
spectral index (i.e., the equivalent of the k-correction) requires
absolute calibrations of near-ultraviolet band-passes.
These calibrations are obtained here by assuming that an
average O3-6 stellar spectrum of 
Fanelli \etal\ (1992\markcite{fanelli92}) approximates a
$4\times 10^4$~K blackbody (Allen 1973\markcite{allen73}) in the 
2400--3000~\AA\ region,
a crude assumption.  Under this assumption, O stars have
$-1.2\la\alpha_{UV}\la0.0$, while F stars have
$6.0\la\alpha_{UV}\la9.8$ and G stars have $\alpha_{UV}\ga9.8$
(\cite{fanelli92}).  Even the galaxies in the sample with
spectral class ``$\cal A$'' have $\alpha_{UV} < 8$ (Figure~6), 
suggesting that they
contain a significant number of main-sequence F stars, which
appears to be consistent with the expected turnoff for
a very old population seen at $z \sim 0.6$ 
(Bertelli \etal\ 1994\markcite{bertelli94}).

An alternative to the near-ultraviolet age dating technique is to
perform a more sophisticated population synthesis based on the entire
observed spectral energy distribution of each galaxy.  In principle
this technique ought to be more sensitive to each galaxy's entire star
formation history, and less sensitive to reddening or the details of
the most recent burst of star formation activity.  On the other hand,
as discussed by 
Van Dokkum \etal\ (1998\markcite{vandokkum98})
it is much more model-dependent, being affected by the details of the
choice of stellar initial mass function, accurate models of stellar
spectra over a broad range of stellar types, and the adopted
possible star formation histories.
Bruzual \& Charlot (1996) compute broad-band colors for single-burst populations as
a function of age; they find a roughly power-law rise in $\alpha_{IR}$
(as defined above in terms of the rest-frame $B$ and $K$ fluxes) from
$\alpha_{IR}\approx -0.2$ at $0.1$~Gyr to $\alpha_{IR}\approx 1.3$
at $10$~Gyr (for a Salpeter mass function, solar metallicity, and
no reddening by dust).  A continuous star formation model rises from
$\alpha_{IR}\approx -0.3$ to $\alpha_{IR}\approx 0.5$ over the same
time period (for the same mass function and metallicity).  A
significant fraction of the galaxies are somewhat redder in $\alpha_{IR}$ than
either model predicts, (Figure~5).  
For example, the ``${\cal A}$'' galaxies in Group 3, which 
are observed at a cosmic
time $t\sim7$~Gyr, have a nominal age based upon their
infrared spectral slope of $\sim10$~Gyr, 
even assuming a single burst of star formation.
This discrepancy between the estimated stellar ages and the 
age of the Universe is a matter of concern, but 
at the very least, it does suggest that 
the ``${\cal A}$'' galaxies were formed in the early Universe
and long before the ``${\cal E}$'' galaxies. 

The UV spectral index $\alpha_{UV}$ becomes bluer as $z$ increases, again in 
all spectral classes, but most prominently in ``${\cal E}$''
galaxies.  This is not unexpected because we know that
there is a pronounced evolution in the OII luminosity 
function (\cite{hogg98oii} and references therein).  However most
of evolution in the OII luminosity density 
is due to the increase in the mean luminosity of
``$\cal{E}$'' galaxies with $z$.

\section{Redshift and Spatial Clustering \label{clustering} }
\subsection{Groups}

As discussed in Cohen \etal\ (1996a\markcite{cohen96a}), and references
cited therein, roughly half the galaxies in deep, pencil beam surveys
out to $z\sim1$ are found in $\sim4-8$ groupings with local velocity
dispersions $\lesssim600$~km s$^{-1}$.  Our re-analysis 
of this field essentially corroborates our earlier results.  
The clustering in redshift space is readily apparent in Figure~1a,b,c.

There are two complementary approaches to analyzing this distribution.
The first, which we shall adopt in a forthcoming analysis of a larger
sample, is to estimate the two point correlation function and its Fourier
transform.  Here, we confine our attention to identifying discrete groups
by smoothing the distribution in local velocity space $V=c\log(1+z)$
with a velocity width of 15,000 km s$^{-1}$
to give the overall distribution in redshift which peaks at $z\sim0.6$. 
We then smooth with a width of 300 km s$^{-1}$ and measure the relative 
overdensity with respect to the overall distribution. Peak overdensities 
in excess of 5 are designated as groups and limited by gaps in the redshift 
distribution (Figure~8). 

\placefigure{fig8}

Although
this procedure is somewhat
subjective, it turns out to be quite instructive.  We identify
five distinct groups, labeled 1-5 in the interval $0<z<0.8$ defined 
by the gaps in the redshift distribution with overdensities
of 5, 10, 13, 7 and 8 respectively.  These 
are the same groups as those we found earlier and are listed in Table~2
and their angular distributions are exhibited in Figures~9. 
We characterize each of the groups in turn.

\placefigure{fig9}

\subsubsection{Group 1.  $0.391\le z\le0.394$}
This group is only marginally overdense
and consists of three ``${\cal E}$'' 
galaxies with similar SEDs concentrated around a 
single, luminous ``${\cal A}$''  red galaxy. 
\subsubsection{Group 2.  $0.428\le z\le0.432$}
This group comprises five galaxies classified spectroscopically as
``${\cal A}$'' and five more classified as ``${\cal C}$'', the former class 
exhibiting steeper UV spectra
than the latter class, as expected.  The luminosities range from $\sim3L_B^\ast$
to $\sim0.1L_B^\ast$, which is close to the survey limit.  There are no 
``${\cal E}$'' galaxies. This is a diffuse group 
with no clear center that may extend beyond our
field.
\subsubsection{Group 3.  $0.577 \le z\le0.588$}

In this group there are 16 galaxies that were classified spectroscopically as ``${\cal A}$'',
one as ``${\cal AC}$''
and 12 galaxies that were classified as ``${\cal C}$''. 
Again, the latter have bluer
UV spectra.  There are several galaxies with redshifts close
to the range we used to define Group 3 which we rejected as 
non-members on the
basis of their being several $\sigma$ separated in
redshift. What is striking is how similar are the shapes of the SEDs,
especially for the ``${\cal A}$'' galaxies (Figure~4c).
Examining the SEDs, we find only one anomaly;
D0K111, classified as ``${\cal C}$'', has a SED more typical of an 
``${\cal A}$''
galaxy.  Ignoring this galaxy, the SEDs
are remarkably uniform in shape within the ``${\cal A}$'' and within the
``${\cal C}$'' members of this peak.

Another interesting point is that there are no  ``${\cal A}$'' galaxies 
in this peak fainter than a minimum luminosity
which is brighter than our survey
cutoff.  This suggests that the ``${\cal A}$'' galaxy luminosity function
falls off at the faint end relative to that of ``${\cal C}$''
and ``${\cal E}$'' galaxies, as it does in the local Universe
(Bingelli, Sandage \&  Tammann 1988\markcite{bingelli88}).

Turning to the sky distribution, we see a ``core'' of four bright galaxies, 
all of which were classified  as spectral class ``{$\cal A$}'',
surrounded by many fainter
galaxies which do not appear to show much central concentration,
except that there are no galaxies in this peak in the top (Northern)
20\% of the field.  In this group, the brightest galaxy at $R$ 
(D0K13) is in fact the ``{$\cal C$}'' galaxy located at
the southern edge of the field, far from the apparent ``core'', but the
most luminous galaxy is D0K08, an ``$\cal{A}$'' galaxy in the central core.

\subsubsection{Group 4.  $0.676\le z\le0.681$}

Here there are seven ``${\cal C}$'' galaxies, one  ``${\cal AC}$'' 
galaxy (which is the one of the two brightest galaxies
in this group) and one ``${\cal E}$'' galaxy surrounding a 
luminous AGN.  There are no ``${\cal A}$'' galaxies.  The UV spectra are correspondingly
bluer than for Group 3.
\subsubsection{Group 5.   $0.761\le z\le0.772$}
This group contains  six ``${\cal A}$'' and three ``${\cal C}$'' galaxies, 
mixed between two separate velocity subgroups.

\subsection{Spectral properties of group galaxies\label{spectra_groups} }

The ``{$\cal E$}'' galaxies avoid the groups while the 
``{$\cal A$}''galaxies are under-represented outside the groups, (Figure~7).
Again this effect has been seen before, particularly in studies
of clusters of galaxies.
Fisher \etal\ (1998) suggest, in their study of a rich galaxy
cluster at $z = 0.33$, that the emission line galaxies have a higher
velocity dispersion than do the absorption line systems and are still 
accreting into the cluster, an idea dating back to the seminal work
of Gunn \& Gott (1972).
Lin \etal\ (1996b\markcite{lin96b}) analyzed the LCRS 
(Shectman \etal\ 1996\markcite{shectman96}) to show that the
non-emission line galaxies show more clustering power over a wide
range of scales than do the emission line galaxies.
In very local surveys,
Salzer (1989) and Rosenberg, Salzer \& Moody (1994) have shown that
emission line galaxies are more isolated (less clustered) than earlier type
galaxies,  
while Benoist \etal\ (1996) claim to detect a strong dependence
of clustering on galaxy luminosity in the SSRS2 sample. 

At higher redshift,
Phillips \etal\ (1997\markcite{phillips97}) studied a sample of 
compact galaxies in the
HDF, most of which are galaxies with strong emission lines and high
rates of star formation.  They suggested that their sample tends to be
less concentrated within the strong redshift peaks in the HDF than is
the total HDF sample, in agreement with what we have found.  This
follows from comparing the redshifts of their HDF sample of blue
compact galaxies to the redshift peaks found in the sample of Cohen
\etal\ (1996b\markcite{cohen96b}).  Deep galaxy counts by Roche \etal \ 
(1996\markcite{roche96}) suggest an angular correlation function
which has higher amplitude for red galaxies than for blue galaxies.
Lacking redshifts, they ascribe this to the much wider redshift range
for faint blue galaxies.

These differences in the behavior of galaxies of various spectral
types, in terms of the shape of their luminosity function, their
maximum luminosity, their tendency to occur in groups, and their
clustering properties, are of great importance in the design and
interpretation of future redshift surveys where sparse sampling
will be utilized.

Figures~4,5,6 show that the independently determined spectroscopic and 
SED classifications for galaxies in our sample are
strongly correlated. The presence of 
emission lines is directly
associated with bluer UV spectra and, indirectly, with bluer IR spectra.
Furthermore there is weaker evidence for a differential effect
within groups.  If we ignore D0K111, 
we find that $\alpha_{UV}$ diminishes as the luminosity 
$L_B$ increases and that $\alpha_{IR}$ becomes redder. Figure~4 shows this
for the $z_p = 0.58$ group.

The ultraviolet extinction for the ``$\cal{E}$'' galaxies was discussed
in \S\ref{uvext}.  The similarity of the spectral indices
among the galaxies in a given group found here
confirms that the variation in UV
extinction among the ``{$\cal A$}'' and ``{$\cal C$}'' 
sample galaxies is a function of spectral type with
small scatter; we suspect that the
reddening itself must be small. 

\subsection{Dynamics of Groups \label{dynamics} }

It appears that the groups are mostly dynamically evolved structures.
All of the five groups show strong evidence of non-uniformity on the sky and
three of the five show an apparent central concentration, suggesting that
some dynamical relaxation has taken place. Group 1, 
with only four members, is too small to make any statement and the field is too small to 
cover all of group 2. We can characterize these groups by their measured 
velocity dispersions (Table~2).
We do this assuming that a 4.7\AA\ instrumental uncertainty 
corresponding to $\sigma(z) {\approx}0.0008$
has been
removed in quadrature.  (The dispersions are slightly smaller than those in 
Table~2 of Cohen \etal\ (1996a\markcite{cohen96a}) due in part to 
re-reduction of one slitmask for which an erroneous wavelength 
scale was used in the 
preliminary reduction, and in part to the removal of the instrumental 
contribution.) 


Note that both groups 3 and 5
show substructure with lower dispersions. In the case of group 3,
a special effort was made to determine $\sigma(v)$ more accurately in
the ``core'' of galaxies in the peak at $z_p = 0.581$.  A template was
prepared by summing up the spectra of the brightest absorption line
galaxies in this peak, and then a cross correlation analysis was
carried out using the wavelength region 5730 to 6896\AA, which covers
3625 to 4360\AA\ in the rest frame.  
$\sigma(v)$ for the four galaxies in the ``core'' is 180~\kms.  This
value fluctuates between 150 and 200~\kms\ as the next three galaxies 
nearest the core are added.\footnote{In this special case, where
all the galaxies are bright with high quality spectra, no observational
error at all was removed.}

If one believes that these are bound systems, and the group crossing times
are $\sim2-4$~Gyr so this is quite likely to be the case,
one can calculate the dynamical masses and hence the mass-to-light ratios.
In the present case, this procedure presents extra difficulties because
some of the groups probably continue off the field. Nonetheless,
if we assume that galaxies are point particles moving
in a dark matter background, then we can define a geometrical and velocity center for
each group and average $r_\perp\Delta V_\parallel^2/G$ to obtain a dynamical
mass estimate (cf Bahcall \& Tremaine 1981\markcite{bahcall81}).
The details of this procedure are 
quite model-dependent, even given complete sampling.  (We choose to assume
that the groups are spatially spherically symmetric, that the velocity distributions
are locally isotropic and that the radial structure is quasi-isothermal.)
On this basis, we compute maximum likelihood 
mass estimates out to the extremities of the observed galaxy 
distribution  for groups 2,3,4,5 given in Table~2.
(Group 1 is too small to carry out this procedure.)
The associated densities in units of the current, mean cosmological density
and the $B$ magnitude mass to light 
ratios are also given in Table~2. 
If we treat group 5 as two 
subgroups, the mass to light ratios are much 
closer to the other three estimates.
These estimates are consonant with values obtained by other methods.
Ramella \etal\ (1997\markcite{ramella97}) obtained a
mean value for their sample of groups of $M/L_B \sim 140$
M\subsun/L\subsun,
while the CNOC survey has measured 295 M\subsun/L\subsun\ (using the
k-corrected R band luminosities; \cite{carlberg96}).  
%
%


Furthermore,
despite the large errors and uncertain cosmological calibration,
it seems reasonable to conclude from the mass density measurement that 
group 3 formed before groups 2 and 4. (The age of the Universe at 
the time of emission are 8.7, 8.4, 7.4, 6.9, 6.5~Gyr for groups
1,2,3,4,5 respectively.) 
If we take the density measurements literally, we conclude that the Universe doubled
in age over the time that groups first formed.  Clearly, a larger database and 
numerical simulation will be necessary to improve upon these rough estimates.

\subsection{Isolated galaxies}

Let us now contrast the 62 galaxies in the five major groups with 
the 45 galaxies
found outside groups with $z<0.8$. It is
apparent that the isolated galaxies are systematically less luminous 
than those in 
groups. KS tests show that the distributions in both observed 
$K$ magnitudes and inferred rest $B$ luminosities are significantly different. 
(Representative luminosity functions are presented below.)
Our results suggest that a luminosity-density relation
exists for galaxies in low density regions.

Also there is a higher proportion of ``${\cal E}$'' galaxies 
in the field. The ratio
of ``${\cal E}$'' to ``${\cal A}$'' galaxies 
is 13 : 8 as opposed to 4 : 26 for the group
galaxies (of which 3 of the 4 ${\cal E}$ galaxies are in 
group 1), with a similar conclusion if we classify the galaxies purely by their SEDs.
In addition, there is an excess of ``${\cal E}$'' galaxies
with $0.1L^\ast\lesssim L\lesssim0.3L^\ast$ with $z\lesssim0.4$.  This may be 
a manifestation of large fluctuations caused by having
only a few groups. 
(The properties of these ``${\cal E}$'' galaxies are in 
contrast to what is observed at high redshift, where
the strong emission line galaxies are very luminous. 
Cowie \etal\ (1996\markcite{cowie96}) first pointed out the change in the behavior with
galaxy luminosity of the 3727\AA\ [\ion{O}{2}] emission; this is also
related to the evolution of the 3727\AA\ [\ion{O}{2}] luminosity function 
with redshift found by Hogg \etal\ (1998\markcite{hogg98oii}) in the HDF.)
Inspecting the sky distribution of these galaxies, we find no strong clustering
beyond occasional pairs.

Most of the apparently isolated galaxies outside the 5 major redshift
peaks are actually members of even smaller structures in redshift space
or of pairs.
The existence of ``isolated'' local field galaxies was discussed by
Vettolani \etal\ (1986\markcite{vettolani86}), who concluded that 
such objects are quite rare.  We reach the same conclusion for
the galaxies in this field at $0.2 < z < 1$.  This can be viewed
as a manifestation of the highly correlated spatial distribution of galaxies.

\subsection{Interpretation of the Groups \label{whatarepeaks} }

Table~2 gives some of the characteristics of the population of each of
the redshift peaks.  The total population of the group, its total luminosity,
the number of galaxies with
$L > L^*$ (calculated at $K$), the value of $L/L^*$ at which the galaxy
(assuming it is an Sa) falls out of a $K < 20$~mag sample, and the
spectral type assigned to the brightest galaxy are given for each of
the five major groups. 

Cohen \etal\ (1996a\markcite{cohen96a}) showed that the strong
redshift peaks do not resemble distant galaxy clusters as rich as
Virgo or richer.  
The velocity dispersion for the early-type galaxies in the relatively
sparse Virgo cluster is 570~\kms\ (\cite{bingelli87}),
significantly larger than that seen in most of the redshift peaks.
The projected surface density of galaxies in the redshift peaks is
much smaller than that in the cores of typical rich clusters, although
it is probably not very different when compared to the peripheral regions.  
Furthermore, the Palomar Deep Cluster Survey (\cite{postman96}) finds only 
seven clusters of galaxies per square degree out to $z\sim0.6$ with richness 
class $\ge1$.
Assuming a radius of 1$h^{-1}$ Mpc for each cluster, we calculate a
probability of $\approx15\%$ of intersecting such a rich cluster for
$z > 0.3$.  Thus the probability of producing multiple redshift peaks
using clusters of galaxies is too small.

Zabludoff \& Mulchaey (1998\markcite{zabludoff98}) have studied nearby poor groups of
galaxies, poor being defined as having $\le5$ members with $L \ge L*$.
These groups are, however, considerably more populous and more massive
than the Local Group, but less concentrated than Hickson's
(1997\markcite{hickson97}) compact groups.  They find groups with 20 -- 50 members down to
$M_B = -15$~mag whose velocity dispersions are 190 to 460~\kms\ and
with typical size of $r_h = 0.5h^{-1}$ Mpc.  Many of these show a
central concentration of early type galaxies.  The latest effort to 
catalog and search for groups in the CfA survey
has been carried out by Ramella \etal\ (1997\markcite{ramella97}).  
The median velocity dispersion of the 406 groups they found is 190~\kms.
Our results show consistency with these findings extending to $z\sim0.8$.

Such groups are much more common than rich clusters of galaxies in the local 
Universe.  
Ramella \etal\ (1997\markcite{ramella97}) found 406 groups, 
149 of which contain 5 or more members, 
in the CfA catalog, while there are only $\sim$15 clusters within this sample 
of richness class 1 or greater.  
Thus the probability that a line of sight to $z \approx 0.8$
will intersect a suitable number of groups so as to produce the
observed redshift peaks seems reasonably high.  Of course this
requires that such groups already exist at $z \sim 0.8$.  
They must be stable old
structures, not ones that collapsed and formed relatively recently.

Group~3 is clearly a structure one might call a poor cluster of galaxies.
The predominance of absorption line galaxies, the large number of members,
the central concentration, and the apparent virialization
contrast strongly with the other looser, less populous redshift peaks 
dominated by galaxies of spectral type ``{\cal C}''
which might properly be called groups.
In Abell's (1958\markcite{abell58}) classification, Group~3, with
$\sim20$ members down to $M_3 + 2$ mag, would 
fall just below his adopted cutoff for richness class 0, the poorest
class of cluster he considers.

How extended perpendicular to the line-of-sight might these strong redshift peaks be?
We have obtained $\approx$500 spectra for a sample of galaxies in fields 
separated by up to 1 degree from this field.
Although the analysis is incomplete, the data
suggest the existence of correlated tructure over
an angular scale of 1 degree (corresponding to a length scale 
of 10 Mpc for $z \sim 0.4$), at least out to $z \sim 0.4$ (\cite{cohen99b}).
However, if this structure is present, it
does not preclude essentially isolated and 
autonomous, virialized groups and clusters. 
Indeed it would be very surprising, on theoretical grounds, if large, 
sheet-like structures did not fragment in this manner (\cite{bertschinger85}). 

We have used this much larger sample  to estimate the frequency of 
``poor clusters'', and find it reasonably close to that obtained
by extrapolating to less populous clusters
the frequency of Abell richness class 1 clusters.
Further discussion of this is deferred to a future paper.

We now turn to the spectroscopy.
Roughly 30\% of the brightest galaxies in
the peaks have been assigned a spectral type of ``$\cal C$'' rather than of
``$\cal A$'', and the spectral class of the brightest galaxy itself
switches from ``$\cal A$'' to ``$\cal C$'' (see Table~2) 
in the higher redshift
peaks.  This may well be a manifestation in galaxies in the
field or groups of the Butcher-Oemler
effect, which was initially associated with galaxy colors.  Butcher \&
Oemler (1984\markcite{butcher84}) found that the bright galaxies in clusters of galaxies
at $z \approx 0.4$ have much bluer colors than normal elliptical
galaxies, with the fraction of blue galaxies reaching 80\% at $z \sim
0.9$ (\cite{rakos95}).  The ``blue galaxies'' have been
studied spectroscopically in rich galaxy clusters by, e.g., Lavery \&
Henry (1988\markcite{lavery88}), Dressler \& Gunn (1992\markcite{dressler92}), 
Caldwell \& Rose (1997\markcite{caldwell97}), and their spectra of such 
objects in rich galaxy clusters look much like ours for field galaxies.

We hypothesize, perhaps simplistically, that
the differences among the redshift peaks are due purely to the amplitude of
the initial perturbation, that this amplitude for the
proto-Group~3 was larger than for any of the other redshift structures, 
and that this in itself produced a structure that collapsed earlier
with stronger virialization and more rapid exhaustion of gas, 
thus leading to the 
differences we see today among the redshift peaks.  In our view there is
no sudden change in any initial property that marks a boundary
between a poor cluster and a rich group, although their 
appearance at the current epoch is quite different.

We have marshalled the evidence which shows quite clearly that,
(for $z\lesssim0.8$), the group galaxies are more luminous and have weaker
UV emission than their counterparts in the field. The striking similarities
of the SEDs within a group leads us to hypothesize that
the galaxies within a group are coeval although different groups did not form simultaneously.  
The groups are then somewhat 
analogous to globular clusters where one can observe the contemporaneous evolution
of stars of different mass.
It is then, in turn, reasonable to suppose that the extreme 
paucity of ``${\cal E}$''
galaxies in these groups (excepting Group 1) is a reflection of their age 
and is analogous
to the presence of a main sequence turnoff in a globular cluster.  On this 
basis, we argue that, at the emission epoch, group 3 is the oldest structure,
followed by groups  5, 2, 4, and 1 in that order. It is re-assuring that this order
is consonant with that obtained on dynamical grounds.   

It appears that these galaxy groups are
higher density regions where the morphology-density relation 
(\cite{dressler80}; \cite{dressler97}),
extended to even lower galaxy density than the rich clusters
where it was originally found, biases the morphological mix 
towards ellipticals and S0 galaxies.
Furthermore our results strongly suggest that high luminosity
galaxies, which presumably are high mass galaxies, form most
readily and almost exclusively in regions of higher galaxy density.
As a group ages, the incidence of star formation within its constituent 
galaxies diminishes. However,
it is necessary to understand the stellar evolution chronometrically
in order to decide whether or not the star formation rate declines secularly or 
is intermittent with decreasing duty cycle. 
  
\section{Evolution of the Luminosity Function and SEDs \label{evolutionoflstar} }
\subsection{Luminosity Functions for $z<0.8$}

We have argued that our sample is quite complete and with a uniform
selection criterion $K<20$, out to $z<0.8$, at least.  However, we have
uncovered dramatic inhomogeneity in the galaxy types. As this variation
is associated with comparatively few structures along a single pencil
beam, we are subject to strong Poissonian fluctuations
and it is not possible to explore evolution in detail using this 
sample alone.  In spite of this, we can use the data set to 
compute volume-averaged luminosity functions out to $z=0.8$. 

Our procedure is to split the data into a few subsets and to fit
Schechter luminosity functions to the shape of these distributions
for each subset, solving for the conventional parameters
$\alpha, L^\ast$ using maximum likelihood estimators (cf Hogg 1998). 
We then used the observed galaxies to normalize the distributions
to obtain estimators of $\Phi^\ast$. After some experimentation,
we find that the minimum number of subsets that is necessary is three;
isolated galaxies, ``${\cal A}$'' galaxies in groups and ``${\cal C}$'' galaxies
(plus the single ``${\cal E}$'' galaxy in a group)
in groups. (Group 1 is excluded.)
(Additional partitioning of the sample does not produce
any more significant results and dividing the galaxies on the basis
of their SEDs rather than their spectroscopy gives similar
conclusions as might be expected.)   This procedure takes into
account the selection at $K$ of the sample, the correlation
between $L_K$ and $\alpha_{IR}$ shown in Figure~6, and the 
differences among the subsets used of their luminosity functions.
The results for $L_B^{\ast}$
constant with $z$ are given in Table~3
and the composite luminosity function is shown in Figure~10.
Calculations with $L_B^{\ast}$ increasing slightly with $z$ show a 
strong covariance between $\alpha$ and $L^{\ast}$.  A reasonable
fit can be obtained with $L_B^{\ast}(z) \propto 10^{0.3z}L_B^{\ast}(z=0)$, but
much larger increases in $L_B^{\ast}$ can be ruled out.  For that
evolution of $L_B^{\ast}$, which corresponds well with passive
evolution in some galaxy models, we find (see Table~3)
$L_B^{\ast}(z=0) = 1.00$ and $\alpha = -1.47$ for the entire sample 
of 108 galaxies out to $z<0.8$, a good match with local values.

\placefigure{fig13}

The two most striking features of this analysis are the smaller value of
$L^\ast$ for the isolated galaxies and the larger value of 
$\alpha$ (flatter luminosity function at the faint end)
for the ``${\cal A}$'' galaxies in the groups. 
Both conclusions are apparent from Figures~1 and 5,
the former from the shortage of $\gtrsim L^\ast$ galaxies outside of groups
and the latter from the paucity of group galaxies close to the selection limit.
Again, neither result is without precedent.  We know from the work of 
Ellis (1997) and references cited therein that there is a relatively local population
of low luminosity galaxies. What we are demonstrating is that
these are found preferentially in low density regions.  Furthermore, 
Bingelli, Sandage \& Tammann (1988\markcite{bingelli88})
argued that, in clusters, the traditional Schechter form with $\alpha\sim-1$
was actually a superposition of luminosity functions 
each more strongly localized in
$L$.   A similar result has also obtained from the LCRS, a sample
dominated by field galaxies, by 
Lin \etal\ (1996a\markcite{lin96a}).
They demonstrate that the faint end of the LF in their sample is
dominated by emission line galaxies.

We have found that this is also true of lower density groups, and have
extended the regime over which this effect is seen out to $z \sim 0.8$.

\subsection{Density Evolution}

With this prescription in hand we can now use the ``low'' 
redshift luminosity functions
to predict the number of galaxies expected in the 
redshift interval $0.8<z<1.3$
on the hypothesis that there is no density evolution. 
These results are given in Table~3 and are
not very sensitive to the assumed cosmological model, as
changing the run of comoving volume with redshift invariably
leads to a compensatory change in the luminosity.
We find that in the no evolution model, 43 galaxies are predicted to have 
$0.8<z<1.3$ and 69 galaxies are expected under the passive evolution
model, whereas, excluding two AGN, only 24 galaxies have redshifts
in this interval.
We hypothesize that a large fraction of the 32 galaxies in the 
galaxy sample without redshifts make up the deficit
(see \S\ref{nozgalaxies})
and that no strong
evolution in cosmological density is required. We see no
evidence for extensive merging among the members of this sample out
to $z \le 1.3$.  

We use the SEDs of the galaxies without redshifts to
support our assertion of \S\ref{nozgalaxies}
that most of the objects without redshifts
are galaxies with $0.8 < z < 1.3$.  Figure~11 shows the
rest frame SEDs for the selection of these  assuming $z = 1$, with the changes
in luminosity and flux as $z$ varies from 0.5 to 1.5 indicated
by the tilted scaled line.  It can be seen that, in many cases, 
the choice of $z \sim 1$ leads to plausible SEDs.
Conversely, these galaxies would 
have unprecedented SEDs at much lower redshifts and would 
be far more luminous than any galaxy population for which we have evidence if their redshifts
were $z\gtrsim2$.

Our hypothesis advanced in \S\ref{nozgalaxies}
that many of the galaxies without measured redshifts
are at $z\sim1$ is crucial to our statements regarding the
absence of mergers.  It is at variance with the conclusion of Kauffmann 
\& Charlot (1998\markcite{kauffmann98}), based upon analysis 
of $K$-selected redshift surveys to brighter limiting magnitude.
It may be possible to test our hypothesis 
in a few cases by integrating longer on a few of these galaxies.
Also relevant in setting constraints on the merger rate is an analysis
of the very close pairs, an issue we defer to a future paper.

\placefigure{fig14}

\section{Galaxy Morphology \label{morphology} }

Morphological classifications on the basis of HST images have been
provided for some of the galaxies in this sample by Driver, Windhorst,
\& Griffiths (1995\markcite{driver95}).  They classify 25 objects in this field as
elliptical galaxies, various types of spirals or irregular galaxies.
Nineteen galaxies are in common between the two samples; the remaining
six are just outside the spatial boundary of our sample.  Although
Driver \etal\ state that ``no attempt has been made to separate stars
from compact ellipticals'', it is interesting to note that both of the
objects in common classified on the basis of morphology as elliptical
galaxies are spectroscopically confirmed galactic stars.  Examination
of the HST images from the STScI archive shows that both of these
objects appear stellar.

The latest type galaxy in common (D0K105)
was classified by Driver \etal\ as an
irregular.  Its spectrum shows very strong
emission at 3727, and also shows emission at 4959\AA, H$\beta$ and
5007\AA.  Ignoring the stars, the correlation between the
morphological classes and the spectroscopic classes for the galaxies
in common is reasonable.  More precise spectral classes for the
galaxies and/or a larger sample with spectroscopy, SEDs, and morphological
classifications are required to strengthen
this statement.

In general, we expect these spectral classes and SEDs to be strongly correlated
with galaxy morphology and this will be testable using a
similar survey in the Hubble Deep Field, currently in 
preparation (\cite{cohen99a}).

\section{Discussion \label{summary} }

A deep galaxy redshift survey has been performed on a single 15 sq
arcmin field with the Keck Telescope. Our sample is carefully selected
at infrared wavelengths where the luminosity should be a good tracer of
the stellar mass (Tegmark \& Peebles 1998\markcite{tegmark98}) 
and the cosmological
corrections to the spectra should be minimized.  To a magnitude limit of
$K=20$, the sample contains 195 objects, including 24
spectroscopically confirmed Galactic
stars. The remaining 171 objects appear to be galaxies and redshifts
have been measured for 139 of these. 109 of these galaxies have
redshifts $z<0.8$ and we argue on observational grounds that our survey
is more than 90 percent complete to this redshift. The remaining 30
redshifts lie in the interval $0.8<z<1.44$ and contain a higher
proportion of uncertain values.  The spectra of the 139 galaxies with
redshifts have been distributed into four classes, ``${\cal E}$''
(emission), ``${\cal C}$'' (composite), ``${\cal A}$'' (absorption) and 
``${\cal Q}$'' (AGN).  In addition we compute rest frame SEDs using our six
band ($UBVRIK$) photometry.  The galaxy redshift distribution shows that
60 percent of the galaxies are found in five compact groups in velocity
space, the largest of which has 29 members.

We have discovered three strong correlations:
\begin{enumerate}
\item The SEDs are closely related to the spectral type 
assigned on the basis of the presence or absence of key diagnostic
features 
([OII]3727, H+K, [OIII]5007, etc), in the
sense that ``${\cal A}$'' galaxies have redder UV and IR
spectra,
whereas ``${\cal E}$'' galaxies have bluer spectra. The association of
a hard ultraviolet continuum with emission lines is unsurprising,
(although it does limit the presence of dust), but the correlation
of the infrared slope is of interest and implies that galaxies 
are more similar in (rest frame)
optical luminosity than infrared luminosity, even when
selected at infrared wavelengths.

This regularity in the behavior of galaxy SEDs, extending over
all galaxy spectral types, across our broad wavelength coverage
and up to $z \sim 1$  
is one reason why photometric redshift techniques
work reasonably
well at least out to $z \sim 1$ for high precision photometric data sets.

\item The spectra and SEDs are correlated with 
galaxy luminosity, most strikingly with the infrared luminosity. 
The ``${\cal A}$'' galaxies are more luminous than the ``${\cal E}$'' galaxies
and, consequently, have redder infrared spectra. 

\item The luminous, red, ``${\cal A}$'' galaxies exhibit a strong preference
for the groups; their ``${\cal E}$'' counterparts inhabit the isolated
regions between groups.  
\end{enumerate}

In many respects, these luminosity-density-spectrum correlations 
extend results 
for rich clusters of galaxies to lower density environments.

In addition to discovering these correlations, we confirm that 
the global luminosity function out to $z\sim0.8$,
at least, exhibits no more than mild luminosity evolution and only 
modest density evolution and is
consistent with the Schechter form.  If we assume passive evolution,
consistent with $\sim1$ magnitude dimming since $z\sim1$, our
derived local luminosity function matches the standard one
in shape and in the derived value of $L^{\ast}$.
However, it appears to be a sum of separate
and distinguishable components with different shapes. The 
galaxies in the
low density regions (which are bluer and more frequently ``${\cal E}$'' type)
have steeper low luminosity slopes; the luminosity function
of the redder ``${\cal A}$'' group galaxies has a much flatter 
low luminosity slope. 

Turning to the $0.8<z<1.3$ interval, we verify that the 
``${\cal E}$'' galaxies
are more luminous than their low redshift counterparts. 
Apart from this, and
taking into account the spectroscopic selection effects, 
there is no evidence 
for strong evolution in the galaxies in our sample. 
Although we have found fewer galaxies 
in this redshift range than we expect by projecting the 
low z luminosity function backward in time, we argue, 
on quantitative and spectral grounds, that 
this deficit is made up by a large fraction of the 32 
galaxies in our galaxy sample for which redshifts were not assigned.
In particular, the very red galaxies, most of which are 
do not have measured redshifts, are probably
the progenitors of the local high luminosity ellipticals. 
If so, they must already be quite old at these epochs.

In order to interpret these findings, we turn first to the
groups. Although 
many of these features may have a much larger lateral coherence than
the field width, it is apparent that they are typically inhomogeneous 
on smaller scales.  A crude estimate of the associated masses 
furnishes core densities ranging from $\sim200-500$ times the current
cosmological density (assuming our cosmology.) The apparent 
crossing times
are sufficiently short compared with the age of the Universe that 
these structures ought to be partially relaxed dynamically, even at
redshifts
$z\sim0.6$. Although the chronology is uncertain,
we argue that our richest and densest group (Group 3), which
might properly be called a ``poor cluster'', separated
out from  the overall expansion of the Universe first and on dynamical
grounds should
contain the oldest galaxies. This is consistent with an analysis of the 
stellar evolutionary history of the SEDs which argues for the presence
of stars with
ages $\sim10$~Gyr in the reddest galaxies.  (Our color-redshift
distributions appear
to be quite inconsistent with some published evolutionary models, 
especially those that postulate an active star formation history.)

We are only able to trace clustering out to $z\sim0.8$, but the 
distribution on the sky of higher 
redshift galaxies found by Steidel \etal\ (1998\markcite{steidel98})
suggests that it originated at earlier times. The
isolated galaxies in low density regions show evidence for recent star
formation.  It is hard to decide 
whether or not these galaxies are genuinely young, or rejuvenated by
recent mergers (e.g. Babul \& Rees 1992\markcite{babul92}). 
On the one hand, the bluer infrared continuum slopes argue for a deficit of 
old stars; on the other, the low luminosities of 
the ``${\cal E}$'' galaxies mean that they
are more susceptible to large spectral changes when they interact with
modest-sized
companions.  We do note, though, that our sample, like others that
preceded it, shows
scant evidence for merging of mature galaxies being a large factor in
global galaxy evolution. 

In this paper, we have reported upon some surprisingly strong
correlations in the 
observed properties of galaxies.  If these are confirmed in other fields
and 
shown to be part of a larger pattern by morphological studies and
investigations
of large scale structure, then there should be optimism that our
understanding of 
the evolution of galaxies will, one day, be placed on as firm a 
physical foundation as the corresponding theory of stars.

%
%

\acknowledgements The entire Keck/LRIS user community owes a huge debt
to Jerry Nelson, Gerry Smith, Bev Oke, and many other people who have
worked to make the Keck Telescope and LRIS a reality.  We are grateful
to the W. M. Keck Foundation, and particularly its late president,
Howard Keck, for the vision to fund the construction of the W. M. Keck
Observatory.  We thank the referee and Greg Bothun for useful suggestions and
Stephane Charlot for providing results of
population synthesis models in advance of publication.
JGC is grateful for partial support from STScI/NASA grant AR-06337.12-94A.
RDB acknowledges support under NSF grant AST95-29170.
DWH and MAP were supported in part by Hubble Fellowship grants 
HF-01093.01-97A and HF-01099.01-97A from STScI (which is operated 
by AURA under NASA contract NAS5-26555).

\clearpage

%
%
\begin{deluxetable}{rrrrrrrrr}
\small
\tablenum{1}
\tiny
\tablecolumns{9}
\tablewidth{0pc}
\tablecaption{Rest Frame SEDs for Galaxies\tablenotemark{a}
}
\label{tab1}
\tablehead{
  \colhead{ID} &
  \colhead{$[{\nu}F_{\nu}]_K$} &
  \colhead{$[{\nu}F_{\nu}]_I$} &
  \colhead{$[{\nu}F_{\nu}]_R$} &
  \colhead{$[{\nu}F_{\nu}]_V$} &
  \colhead{$[{\nu}F_{\nu}]_B$}\tablenotemark{b} &
  \colhead{$[{\nu}F_{\nu}]_U$} &
  \colhead{$[{\nu}F_{\nu}]_P$} &
  \colhead{$[{\nu}F_{\nu}]_Q$} 
\\
Log($\nu$) &
(14.14) &
(14.54) &
(14.66) &
(14.74) &
(14.83) &
(14.91) &
(15.00) &
(15.10) 
}
\startdata 
  18 &  $-$0.69  &  $-$0.49  &  $-$0.46  &  $-$0.47  &  $-$0.49  &  $-$0.74  &  $-$1.02  &  $-$1.34  \nl 
 104 &  $-$1.48  &  $-$1.21  &  $-$1.14  &  $-$1.11  &  $-$1.17  &  $-$1.20  &  $-$1.24  &  $-$1.27  \nl 
  88 &  $-$1.34  &  $-$1.09  &  $-$1.04  &  $-$1.04  &  $-$1.06  &  $-$1.29  &  $-$1.62  &  $-$1.98  \nl 
  54 &  $-$0.92  &  $-$0.91  &  $-$0.94  &  $-$0..99  &  $-$0.95  &  $-$1.14  &  $-$1.47  &  $-$1.83  \nl 
  71 &  $-$1.15  &  $-$0.90  &  $-$0.86  &  $-$0.87  &  $-$0.90  &  $-$1.08  &  $-$1.33  &  $-$1.62  \nl 
 149 &  $-$1.32  &  $-$1.04  &  $-$0.96  &  $-$0.91  &  $-$0.68  &  $-$0.99  &  $-$1.13  &  $-$1.29  \nl 
  25 &  $-$0.31  &  $-$0.16  &  $-$0.12  &  $-$0.11  &  $-$0.09  &  $-$0.26  &  $-$0.45  &  $-$0.64  \nl 
 187 &  $-$1.32  &  $-$0.93  &  $-$0.81  &  $-$0.83  &  $-$0.86  &  $-$0.98  &  $-$1.15  &  $-$1.28  \nl 
 171 &  $-$1.16  &  $-$0.99  &  $-$0.94  &  $-$0.94  &  $-$0.94  &  $-$1.14  &  $-$1.35  &  $-$1.58  \nl 
  11 &   0.23  &  $-$0.07  &  $-$0.16  &  $-$0.22  &  $-$0.28  &  $-$0.62  &  $-$1.20  &  $-$1.91  \nl 
  57 &  $-$0.46  &  $-$0.43  &  $-$0.42  &  $-$0.42  &  $-$0.43  &  $-$0.54  &  $-$0.68  &  $-$0.70  \nl 
 144 &  $-$0.95  &  $-$0.91  &  $-$0.90  &  $-$0.89  &  $-$0.89  &  $-$1.02  &  $-$1.15  &  $-$1.32  \nl 
 110 &  $-$0.85  &  $-$0.54  &  $-$0.45  &  $-$0.42  &  $-$0.41  &  $-$0.52  &  $-$0.66  &  $-$0.83  \nl 
 132 &  $-$0.94  &  $-$0.60  &  $-$0.50  &  $-$0.47  &  $-$0.46  &  $-$0.57  &  $-$0.73  &  $-$0.75  \nl 
  17 &   0.16  &   0.09  &   0.06  &   0.01  &  $-$0.08  &  $-$0.48  &  $-$1.02  &  $-$1.88  \nl 
 159 &  $-$1.01  &  $-$0.81  &  $-$0.75  &  $-$0.71  &  $-$0.65  &  $-$0.69  &  $-$0.76  &  $-$0.74  \nl 
  81 &  $-$0.59  &  $-$0.40  &  $-$0.34  &  $-$0.32  &  $-$0.29  &  $-$0.47  &  $-$0.55  &  $-$0.78  \nl 
 147 &  $-$0.80  &  $-$0.88  &  $-$0.90  &  $-$0.96  &  $-$1.07  &  $-$1.13  &  $-$1.42  &  $-$1.43  \nl 
   6 &   0.81  &   0.62  &   0.57  &   0.50  &   0.41  &  $-$0.02  &  $-$0.47  &  $-$1.43  \nl 
  14 &   0.31  &   0.19  &   0.15  &   0.09  &  $-$0.01  &  $-$0.39  &  $-$0.95  &  $-$1.64  \nl 
  35 &  $-$0.02  &  $-$0.02  &  $-$0.02  &  $-$0.06  &  $-$0.14  &  $-$0.39  &  $-$0.75  &  $-$1.07  \nl 
 138 &  $-$0.78  &  $-$0.73  &  $-$0.71  &  $-$0.74  &  $-$0.81  &  $-$1.02  &  $-$1.33  &  $-$1.61  \nl 
  33 &   0.05  &  $-$0.10  &  $-$0.15  &  $-$0.22  &  $-$0.33  &  $-$0.71  &  $-$1.15  &  $-$1.64  \nl 
  83 &  $-$0.45  &  $-$0.60  &  $-$0.65  &  $-$0.72  &  $-$0.85  &  $-$1.03  &  $-$1.36  &  $-$1.79  \nl 
  31 &   0.03  &  $-$0.06  &  $-$0.08  &  $-$0.14  &  $-$0.24  &  $-$0.60  &  $-$1.11  &  $-$1.92  \nl 
  16 &   0.28  &   0.14  &   0.10  &   0.06  &  $-$0.01  &  $-$0.31  &  $-$0.63  &  $-$1.01  \nl 
  55 &  $-$0.21  &  $-$0.26  &  $-$0.28  &  $-$0.32  &  $-$0.39  &  $-$0.59  &  $-$0.88  &  $-$1.17  \nl 
 161 &  $-$0.85  &  $-$0.88  &  $-$0.89  &  $-$0.92  &  $-$0.98  &  $-$1.19  &  $-$1.35  &  $-$1.65  \nl 
  87 &  $-$0.53  &  $-$0.42  &  $-$0.39  &  $-$0.38  &  $-$0.37  &  $-$0.50  &  $-$0.63  &  $-$0.68  \nl 
   9 &   0.61  &   0.39  &   0.32  &   0.26  &   0.16  &  $-$0.14  &  $-$0.49  &  $-$0.94  \nl 
 167 &  $-$0.96  &  $-$0.70  &  $-$0.63  &  $-$0.62  &  $-$0.66  &  $-$0.79  &  $-$0.95  &  $-$1.05  \nl 
 102 &  $-$0.43  &  $-$0.74  &  $-$0.83  &  $-$0.95  &  $-$1.15  &  $-$1.21  &  $-$1.31  &  $-$1.42  \nl 
 189 &  $-$0.68  &  $-$1.07  &  $-$1.19  &  $-$1.30  &  $-$1.36  &  $-$1.65  &  $-$1.84  &  $-$2.05  \nl 
 151 &  $-$0.65  &  $-$0.70  &  $-$0.71  &  $-$0.75  &  $-$0.74  &  $-$1.06  &  $-$1.25  &  $-$1.46  \nl 
  73 &  $-$0.39  &  $-$0.14  &  $-$0.06  &  $-$0.03  &   0.04  &  $-$0.17  &  $-$0.36  &  $-$0.38  \nl 
 153 &  $-$0.70  &  $-$0.67  &  $-$0.66  &  $-$0.67  &  $-$0.65  &  $-$0.85  &  $-$1.09  &  $-$1.21  \nl 
  58 &  $-$0.16  &  $-$0.10  &  $-$0.08  &  $-$0.08  &  $-$0.06  &  $-$0.31  &  $-$0.49  &  $-$0.69  \nl 
  90 &  $-$0.32  &  $-$0.40  &  $-$0.43  &  $-$0.46  &  $-$0.46  &  $-$0.64  &  $-$0.78  &  $-$0.79  \nl 
  97 &  $-$0.15  &   0.00  &   0.59  &  $-$0.72  &  $-$0.82  &  $-$0.91  &  $-$1.04  &  $-$0.99  \nl 
  79 &  $-$0.23  &  $-$0.27  &  $-$0.28  &  $-$0.31  &  $-$0.30  &  $-$0.64  &  $-$0.99  &  $-$1.11  \nl 
 170 &  $-$0.82  &  $-$0.51  &  $-$0.42  &  $-$0.37  &  $-$0.29  &  $-$0.49  &  $-$0.62  &  $-$0.69  \nl 
  63 &  $-$0.05  &  $-$0.17  &  $-$0.21  &  $-$0.25  &  $-$0.26  &  $-$0.52  &  $-$0.69  &  $-$0.82  \nl 
 157 &  $-$0.69  &  $-$0.55  &  $-$0.50  &  $-$0.49  &  $-$0.44  &  $-$0.67  &  $-$0.91  &  $-$0.93  \nl 
  86 &  $-$0.37  &  $-$0.15  &  $-$0.09  &  $-$0.05  &   0.01  &  $-$0.14  &  $-$0.21  &  $-$0.30  \nl 
 127 &  $-$0.62  &  $-$0.28  &  $-$0.18  &  $-$0.12  &  $-$0.03  &  $-$0.21  &  $-$0.21  &  $-$0.25  \nl 
 122 &  $-$0.50  &  $-$0.36  &  $-$0.32  &  $-$0.29  &  $-$0.26  &  $-$0.56  &  $-$0.70  &  $-$0.79  \nl 
 139 &  $-$0.24  &  $-$0.70  &  $-$0.84  &  $-$0.93  &  $-$1.04  &  $-$1.09  &  $-$1.05  &  $-$1.11  \nl 
  27 &   0.44  &   0.22  &   0.15  &   0.11  &   0.06  &  $-$0.32  &  $-$0.63  &  $-$0.98  \nl 
  32 &   0.44  &   0.14  &   0.05  &  $-$0.01  &  $-$0.09  &  $-$0.45  &  $-$0.85  &  $-$1.25  \nl 
  48 &   0.23  &  $-$0.01  &  $-$0.08  &  $-$0.13  &  $-$0.19  &  $-$0.39  &  $-$0.63  &  $-$0.88  \nl 
 101 &  $-$0.25  &  $-$0.34  &  $-$0.36  &  $-$0.38  &  $-$0.41  &  $-$0.76  &  $-$1.04  &  $-$1.31  \nl 
   8 &   0.93  &   0.66  &   0.58  &   0.53  &   0.47  &   0.03  &  $-$0.51  &  $-$1.16  \nl 
 131 &  $-$0.38  &  $-$0.49  &  $-$0.52  &  $-$0.54  &  $-$0.56  &  $-$0.92  &  $-$1.24  &  $-$1.60  \nl 
 163 &  $-$0.52  &  $-$0.65  &  $-$0.68  &  $-$0.71  &  $-$0.74  &  $-$1.15  &  $-$1.66  &  $-$2.24  \nl 
  22 &   0.56  &   0.36  &   0.30  &   0.26  &   0.22  &  $-$0.23  &  $-$0.71  &  $-$1.30  \nl 
  45 &   0.25  &   0.07  &   0.02  &  $-$0.02  &  $-$0.07  &  $-$0.51  &  $-$0.89  &  $-$1.34  \nl 
  70 &  $-$0.02  &  $-$0.19  &  $-$0.24  &  $-$0.27  &  $-$0.31  &  $-$0.68  &  $-$1.06  &  $-$1.48  \nl 
  89 &  $-$0.15  &  $-$0.31  &  $-$0.36  &  $-$0.39  &  $-$0.43  &  $-$0.86  &  $-$1.29  &  $-$1.77  \nl 
 145 &  $-$0.49  &  $-$0.50  &  $-$0.50  &  $-$0.50  &  $-$0.50  &  $-$0.77  &  $-$0.96  &  $-$1.18  \nl 
 165 &  $-$0.52  &  $-$0.65  &  $-$0.68  &  $-$0.71  &  $-$0.74  &  $-$1.20  &  $-$1.70  &  $-$2.27  \nl 
  19 &   0.59  &   0.42  &   0.37  &   0.34  &   0.30  &  $-$0.14  &  $-$0.63  &  $-$1.24  \nl 
  28 &   0.42  &   0.25  &   0.20  &   0.17  &   0.13  &  $-$0.29  &  $-$0.79  &  $-$1.41  \nl 
  29 &   0.44  &   0.24  &   0.18  &   0.14  &   0.10  &  $-$0.30  &  $-$0.81  &  $-$1.44  \nl 
  40 &   0.26  &   0.13  &   0.10  &   0.07  &   0.05  &  $-$0.36  &  $-$0.88  &  $-$1.54  \nl 
  47 &   0.21  &   0.02  &  $-$0.03  &  $-$0.07  &  $-$0.11  &  $-$0.42  &  $-$0.73  &  $-$0.99  \nl 
 111 &  $-$0.29  &  $-$0.41  &  $-$0.45  &  $-$0.47  &  $-$0.50  &  $-$1.00  &  $-$1.61  &  $-$2.29  \nl 
 133 &  $-$0.47  &  $-$0.40  &  $-$0.38  &  $-$0.37  &  $-$0.36  &  $-$0.48  &  $-$0.56  &  $-$0.65  \nl 
  39 &   0.35  &   0.09  &   0.01  &  $-$0.04  &  $-$0.10  &  $-$0.45  &  $-$0.86  &  $-$1.36  \nl 
 114 &  $-$0.39  &  $-$0.34  &  $-$0.33  &  $-$0.32  &  $-$0.31  &  $-$0.51  &  $-$0.73  &  $-$0.86  \nl 

\tablebreak

  49 &   0.15  &   0.04  &   0.01  &  $-$0.01  &  $-$0.03  &  $-$0.30  &  $-$0.60  &  $-$0.75  \nl 
  62 &   0.05  &  $-$0.09  &  $-$0.13  &  $-$0.16  &  $-$0.20  &  $-$0.49  &  $-$0.76  &  $-$0.94  \nl 
  67 &   0.04  &  $-$0.17  &  $-$0.23  &  $-$0.27  &  $-$0.32  &  $-$0.72  &  $-$1.16  &  $-$1.65  \nl 
 116 &  $-$0.30  &  $-$0.46  &  $-$0.51  &  $-$0.54  &  $-$0.58  &  $-$0.91  &  $-$1.27  &  $-$1.67  \nl 
  80 &  $-$0.04  &  $-$0.30  &  $-$0.38  &  $-$0.43  &  $-$0.49  &  $-$0.69  &  $-$0.74  &  $-$1.10  \nl 
 164 &  $-$0.58  &  $-$0.55  &  $-$0.54  &  $-$0.53  &  $-$0.52  &  $-$0.72  &  $-$0.90  &  $-$1.10  \nl 
  13 &   0.68  &   0.51  &   0.45  &   0.42  &   0.38  &   0.14  &   0.03  &  $-$0.18  \nl 
  36 &   0.53  &   0.13  &   0.01  &  $-$0.07  &  $-$0.16  &  $-$0.41  &  $-$0.73  &  $-$1.07  \nl 
 103 &  $-$0.36  &  $-$0.15  &  $-$0.09  &  $-$0.04  &   0.00  &  $-$0.16  &  $-$0.24  &  $-$0.37  \nl 
 166 &  $-$0.57  &  $-$0.49  &  $-$0.47  &  $-$0.45  &  $-$0.44  &  $-$0.58  &  $-$0.72  &  $-$0.79  \nl 
  30 &   0.47  &   0.28  &   0.22  &   0.18  &   0.14  &  $-$0.17  &  $-$0.49  &  $-$0.76  \nl 
 184 &  $-$0.69  &  $-$0.49  &  $-$0.43  &  $-$0.39  &  $-$0.34  &  $-$0.50  &  $-$0.65  &  $-$0.69  \nl 
  85 &  $-$0.10  &  $-$0.15  &  $-$0.17  &  $-$0.18  &  $-$0.19  &  $-$0.38  &  $-$0.51  &  $-$0.62  \nl 
  78 &  $-$0.10  &  $-$0.03  &  $-$0.01  &   0.01  &   0.03  &  $-$0.06  &  $-$0.12  &  $-$0.19  \nl 
 162 &  $-$0.49  &  $-$0.46  &  $-$0.45  &  $-$0.44  &  $-$0.44  &  $-$0.54  &  $-$0.68  &  $-$0.77  \nl 
  44 &   0.37  &   0.21  &   0.16  &   0.13  &   0.09  &  $-$0.15  &  $-$0.37  &  $-$0.57  \nl 
  41 &   0.43  &   0.21  &   0.14  &   0.10  &   0.05  &  $-$0.34  &  $-$0.73  &  $-$1.15  \nl 
  74 &   0.27  &  $-$0.16  &  $-$0.29  &  $-$0.37  &  $-$0.47  &  $-$0.80  &  $-$1.20  &  $-$1.65  \nl 
  51 &   0.39  &   0.12  &   0.04  &  $-$0.01  &  $-$0.08  &  $-$0.44  &  $-$0.77  &  $-$1.13  \nl 
  96 &   0.10  &  $-$0.25  &  $-$0.36  &  $-$0.43  &  $-$0.51  &  $-$0.67  &  $-$0.92  &  $-$1.19  \nl 
  84 &  $-$0.06  &  $-$0.04  &  $-$0.04  &  $-$0.03  &  $-$0.03  &  $-$0.21  &  $-$0.34  &  $-$0.47  \nl 
  24 &   0.77  &   0.38  &   0.26  &   0.18  &   0.10  &  $-$0.19  &  $-$0.58  &  $-$0.92  \nl 
  34 &   0.55  &   0.38  &   0.32  &   0.29  &   0.25  &  $-$0.13  &  $-$0.48  &  $-$0.86  \nl 
  52 &   0.30  &   0.19  &   0.16  &   0.14  &   0.11  &  $-$0.11  &  $-$0.31  &  $-$0.52  \nl 
 100 &   0.09  &  $-$0.29  &  $-$0.40  &  $-$0.48  &  $-$0.56  &  $-$0.84  &  $-$1.13  &  $-$1.46  \nl 
  12 &   0.65  &   0.88  &   0.95  &   1.00  &   1.05  &   1.01  &   0.96  &   1.02  \nl 
 105 &  $-$0.15  &  $-$0.13  &  $-$0.13  &  $-$0.12  &  $-$0.11  &  $-$0.20  &  $-$0.26  &  $-$0.26  \nl 
  56 &   0.41  &   0.01  &  $-$0.11  &  $-$0.19  &  $-$0.29  &  $-$0.55  &  $-$0.71  &  $-$0.69  \nl 
 178 &  $-$0.58  &  $-$0.25  &  $-$0.15  &  $-$0.08  &  $-$0.01  &  $-$0.18  &  $-$0.42  &  $-$0.54  \nl 
 143 &  $-$0.03  &  $-$0.32  &  $-$0.41  &  $-$0.47  &  $-$0.53  &  $-$0.90  &  $-$1.07  &  $-$1.26  \nl 
 142 &  $-$0.04  &  $-$0.28  &  $-$0.35  &  $-$0.40  &  $-$0.46  &  $-$0.73  &  $-$1.05  &  $-$1.40  \nl 
  38 &   0.64  &   0.42  &   0.35  &   0.31  &   0.25  &  $-$0.03  &  $-$0.32  &  $-$0.65  \nl 
  46 &   0.42  &   0.43  &   0.44  &   0.44  &   0.44  &   0.28  &   0.16  &   0.11  \nl 
  75 &   0.17  &   0.16  &   0.15  &   0.15  &   0.15  &  $-$0.03  &  $-$0.15  &  $-$0.24  \nl 
 179 &  $-$0.30  &  $-$0.40  &  $-$0.43  &  $-$0.45  &  $-$0.47  &  $-$0.69  &  $-$0.88  &  $-$1.09  \nl 
  59 &   0.48  &   0.17  &   0.08  &   0.02  &  $-$0.05  &  $-$0.28  &  $-$0.36  &  $-$0.65  \nl 
  66 &   0.58  &   0.00  &  $-$0.17  &  $-$0.29  &  $-$0.42  &  $-$0.63  &  $-$1.01  &  $-$1.44  \nl 
  37 &   0.73  &   0.40  &   0.31  &   0.24  &   0.16  &  $-$0.19  &  $-$0.51  &  $-$0.87  \nl 
 140 &  $-$0.15  &  $-$0.18  &  $-$0.19  &  $-$0.20  &  $-$0.20  &  $-$0.39  &  $-$0.64  &  $-$0.76  \nl 
  60 &   0.78  &   0.01  &  $-$0.22  &  $-$0.37  &  $-$0.55  &  $-$0.85  &  $-$1.13  &  $-$1.45  \nl 
 150 &  $-$0.02  &  $-$0.12  &  $-$0.15  &  $-$0.17  &  $-$0.19  &  $-$0.38  &  $-$0.46  &  $-$0.51  \nl 
 137 &  $-$0.12  &   0.01  &   0.05  &   0.08  &   0.10  &  $-$0.02  &  $-$0.08  &  $-$0.02  \nl 
 183 &  $-$0.14  &  $-$0.18  &  $-$0.19  &  $-$0.20  &  $-$0.22  &  $-$0.35  &  $-$0.50  &  $-$0.41  \nl 
 175 &  $-$0.10  &  $-$0.13  &  $-$0.14  &  $-$0.14  &  $-$0.15  &  $-$0.31  &  $-$0.41  &  $-$0.36  \nl 
 121 &   0.08  &   0.12  &   0.13  &   0.14  &   0.15  &   0.02  &  $-$0.04  &  $-$0.09  \nl 
 174 &   0.37 & $-$0.24 & $-$0.42 & $-$0.54 & $-$0.68 & $-$0.73 & $-$0.79 & $-$0.85 \nl
 172 &  $-$0.02  &  $-$0.06  &  $-$0.07  &  $-$0.08  &  $-$0.09  &  $-$0.15  &  $-$0.29  &  $-$0.52  \nl 
 185 &  $-$0.10  &  $-$0.12  &  $-$0.12  &  $-$0.13  &  $-$0.13  &  $-$0.17  &  $-$0.19  &  $-$0.19  \nl
 507 &   0.25  &   0.23  &   0.22  &   0.22  &   0.23  &   0.08  & $-$0.07 & $-$0.22 \nl 
 125 &   0.28  &   0.13  &   0.09  &   0.06  &   0.03  &  $-$0.12  &  $-$0.25  &  $-$0.34  \nl 
 108 &   0.31  &   0.22  &   0.19  &   0.17  &   0.15  &   0.06  &  $-$0.04  &  $-$0.10  \nl 
 181 &   0.18  &  $-$0.11  &  $-$0.19  &  $-$0.25  &  $-$0.31  &  $-$0.35  &  $-$0.45  &  $-$0.75  \nl 
  93 &   0.29  &   0.39  &   0.42  &   0.44  &   0.50  &   0.40  &   0.23  &  $-$0.09  \nl 
  64 &   1.10  &   0.40  &   0.19  &   0.05  &  $-$0.10  &  $-$0.30  &  $-$0.55  &  $-$0.82  \nl 
 152 &   0.42  &   0.07  &  $-$0.04  &  $-$0.11  &  $-$0.22  &  $-$0.21  &  $-$0.23  &  $-$0.25  \nl 
  26 &   1.26  &   0.96  &   0.87  &   0.81  &   0.75  &   0.69  &   0.64  &   0.60  \nl 
  42 &   1.34  &   0.79  &   0.63  &   0.52  &   0.40  &   0.22  &  $-$0.05  &  $-$0.27  \nl 
 182 &   0.25  &   0.01  &  $-$0.06  &  $-$0.11  &  $-$0.16  &  $-$0.18  &  $-$0.20  &  $-$0.41  \nl 
  76 &   0.75  &   0.55  &   0.49  &   0.45  &   0.40  &   0.28  &   0.03  &  $-$0.12  \nl 
 188 &   0.07  &   0.11  &   0.12  &   0.13  &   0.15  &   0.14  &   0.12  &  $-$0.16  \nl 
 135 &   0.57  &   0.30  &   0.22  &   0.17  &   0.11  &   0.00  &  $-$0.29  &  $-$0.60  \nl 
  65 &   1.28  &   0.65  &   0.46  &   0.33  &   0.19  &   0.06  &  $-$0.10  &  $-$0.29  \nl 
 113 &   1.00  &   0.39  &   0.21  &   0.09  &  $-$0.05  &  $-$0.14  &  $-$0.06  &  $-$0.13  \nl 
 112 &   0.68  &   0.47  &   0.41  &   0.36  &   0.32  &   0.27  &   0.18  &   0.15  \nl 
 126 &   0.45  &   0.44  &   0.43  &   0.43  &   0.39  &   0.42  &   0.42  &   0.37  \nl 
 158 &   0.49  &   0.29  &   0.23  &   0.19  &   0.15  &   0.10  &   0.00  &  $-$0.10  \nl 
 128 &   0.55  &   0.43  &   0.39  &   0.37  &   0.34  &   0.30  &   0.10  &  $-$0.04  \nl 
  69 &   1.23  &   0.75  &   0.61  &   0.51  &   0.40  &   0.31  &   0.12  &  $-$0.10  \nl 
  68 &   1.08  &   0.85  &   0.78  &   0.73  &   0.83  &   0.64  &   0.44  &   0.12  \nl 
 123 &   0.85  &   0.55  &   0.46  &   0.40  &   0.34  &   0.27  &   0.25  &   0.26  \nl 

\enddata

\tablenotetext{a} { {$[{\nu}F_{\nu}]_B$} is normalized to 
$\log L_B^\ast=36.86$~W $\equiv M_B^\ast=-20.8$.}
\tablenotetext{b} {$\alpha_{IR} = 1-(\log L_K-\log L_B)/(\log\nu_B-\log\nu_K)$ and
$\alpha_{UV}=1-(\log L_B-\log L_Q)/(\log\nu_B-\log\nu_Q)$}.

\end{deluxetable}

%
%
\begin{deluxetable}{lrrrccccc}
\tablenum{2}
\tablewidth{0pt}
\scriptsize
\tablecaption{Characteristics of the Redshift Peaks}
\label{tab2}
\tablehead{\colhead{$z$ range} & \colhead{$N_p$} & \colhead{N} &
\colhead{Total $L$} &
\colhead{$L/L*$} & \colhead{spectral type} 
& \colhead{$\sigma_v$\tablenotemark{c}}
& $M/L$ & 
\colhead{$\delta\rho/\rho$} \nl 
\colhead{} & \colhead{} &
\colhead{($L > L^{\ast})_K$} & \colhead{($L^{\ast}$)}   & \colhead{(for $K = 20$ mag)}
& \colhead{of brightest} & \colhead{(\kms)} &
\colhead{$(M_{\odot}/L_{\odot})_B$} & \colhead{(at $z = 0$)} \nl
\colhead{} & \colhead{} & \colhead{} & \colhead{} & \colhead{} &
\colhead{galaxy} & \colhead{(Rest Frame)} 
}
\startdata
0.391 -- 0.394 &  4 & 0 & 2 & 0.05 & $\cal{A}$  &  250 & ... & ... \nl
0.428 -- 0.432 & 10 & 1 & 7 & 0.06 & $\cal{A}$  & 300 & 300 & 250 \nl
0.577 -- 0.588 & 29 & 8 & 25 & 0.12 & $\cal{A}$ & 495 & 165 & 665 \nl
0.676 -- 0.681 & 10 & 4\tablenotemark{a} & 19 & 
     0.16 & $\cal{C}$\tablenotemark{b} & 260 & 100 & 180 \nl
0.761 -- 0.772 & 9 & 4 & 10 & 0.21 & $\cal{C}$ & 785 & 1500 & 480 \nl
\enddata
\tablenotetext{a}{Includes one QSO}
\tablenotetext{b}{Excluding the QSO}
\tablenotetext{c}{An observational error of 175~\kms\ has
been removed in quadrature.}
\end{deluxetable}


%
%
\begin{deluxetable}{lrrrrc}
\tablenum{3}
\tablewidth{0pt}
\scriptsize
\tablecaption{Parameters of the Fit Luminosity Functions For $z < 0.8$}
\label{tab3}
\tablehead{\colhead{Galaxy Set} & \colhead{$N_{gal}$} & 
\colhead{$L_B^{\ast}$} &
\colhead{$\alpha$} & \colhead{$\phi^{\ast}$} & 
\colhead{$N_{pred}$} \nl
\colhead{} & \colhead{} &
\colhead{($L_B^{\ast}$ at $z=0)$}  & \colhead{}
& \colhead{($Mpc^{-3}$)} & 
\colhead{($0.8 < z < 1.3$)}
}
\startdata
No evolution\tablenotemark{a} \nl
Isolated & 51 & 0.96 & $-$1.75 & 2.93 & 11 \nl
Group $\cal{C}$\tablenotemark{b} & 33 & 1.34 & $-$1.32 & 2.11 & 14 \nl
Group $\cal{A}$ & 24 & 0.96 & $-$0.34 & 3.31 & 18 \nl
All & 108 & 1.61 & $-$1.52 & 4.71 & 42 \nl
Passive evolution\tablenotemark{c} \nl
Isolated & 51 & 0.63 & $-$1.71 & 5.33 & 8 \nl
Group $\cal{C}$\tablenotemark{b} & 33 & 0.78 & $-$1.25 & 3.04 & 10 \nl
Group $\cal{A}$ & 24 & 0.66 & $-$0.36 & 5.45 & 19 \nl
All & 108 & 1.00 & $-$1.47 & 5.62 & 69 \nl

\enddata
\tablenotetext{a}{$L^{\ast}$ is assumed to be constant to $z = 0.8$.}
\tablenotetext{b}{Includes one ``$\cal{E}$'' galaxy as well.}
\tablenotetext{c}{$L^{\ast}(z) \propto 10^{0.3z}L^{\ast}(z=0)$.}
\end{deluxetable}

\clearpage



\clearpage

%
%

\begin{figure}
\epsscale{0.35}
\plottwo{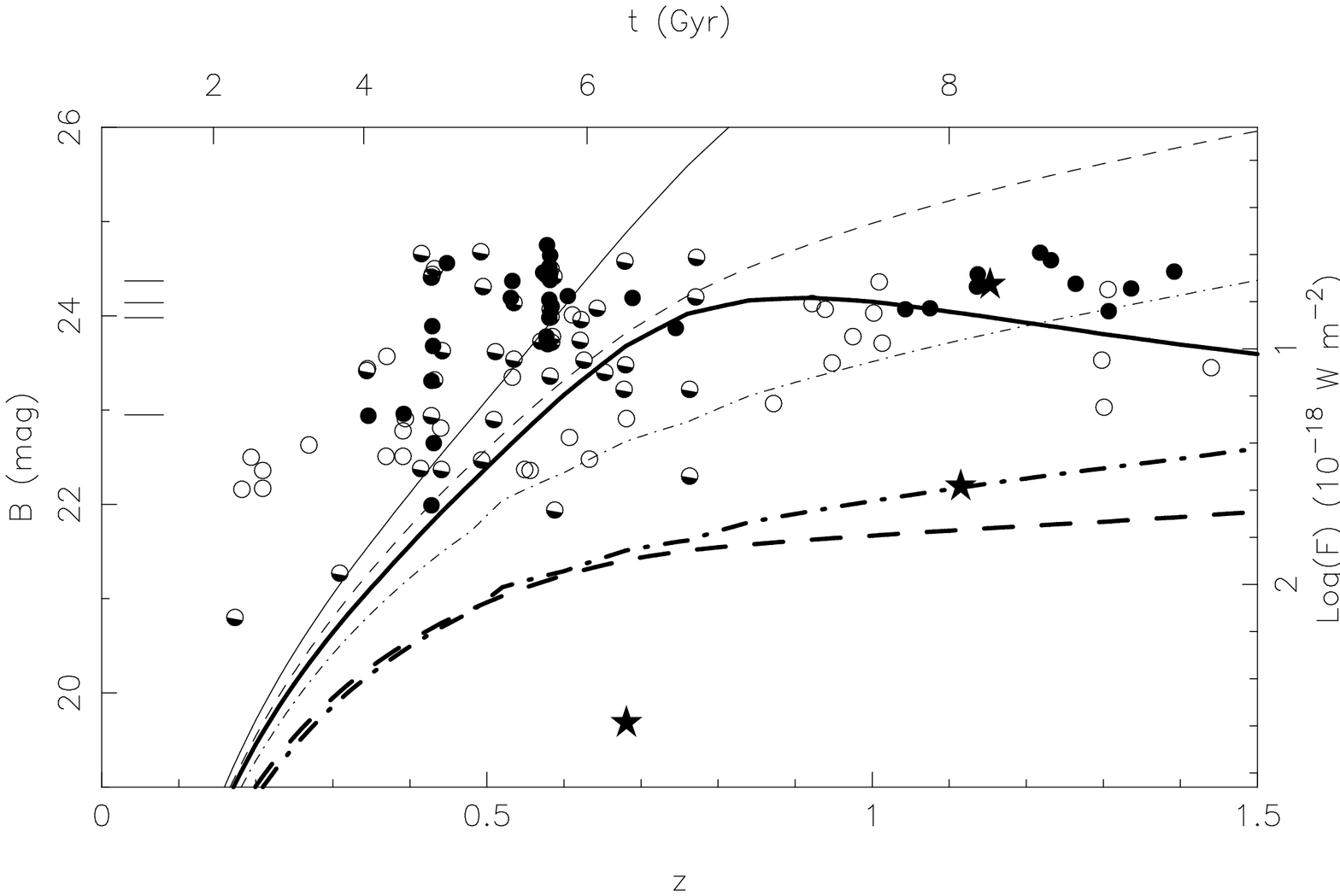}{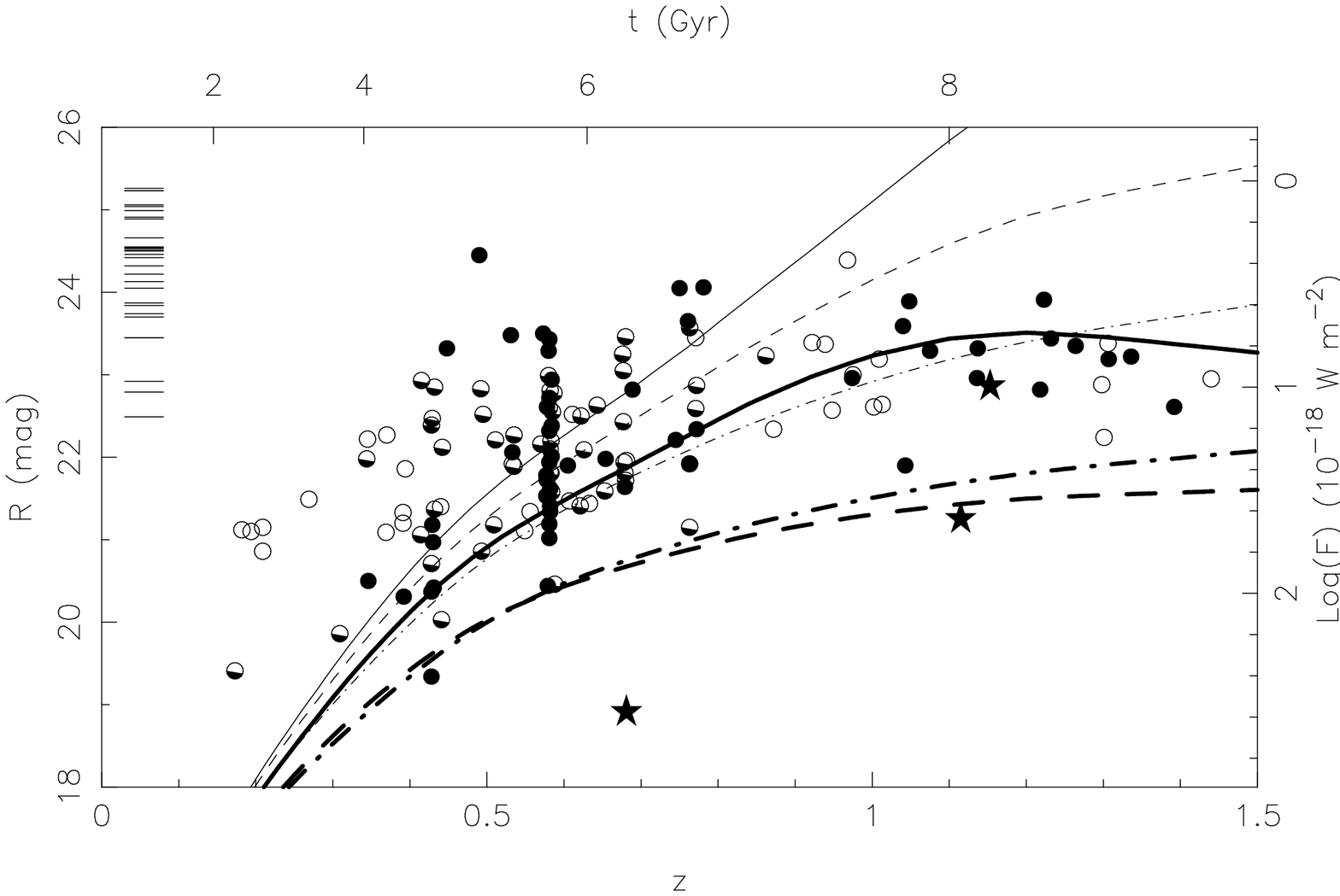}
\epsscale{0.70}
\plotone{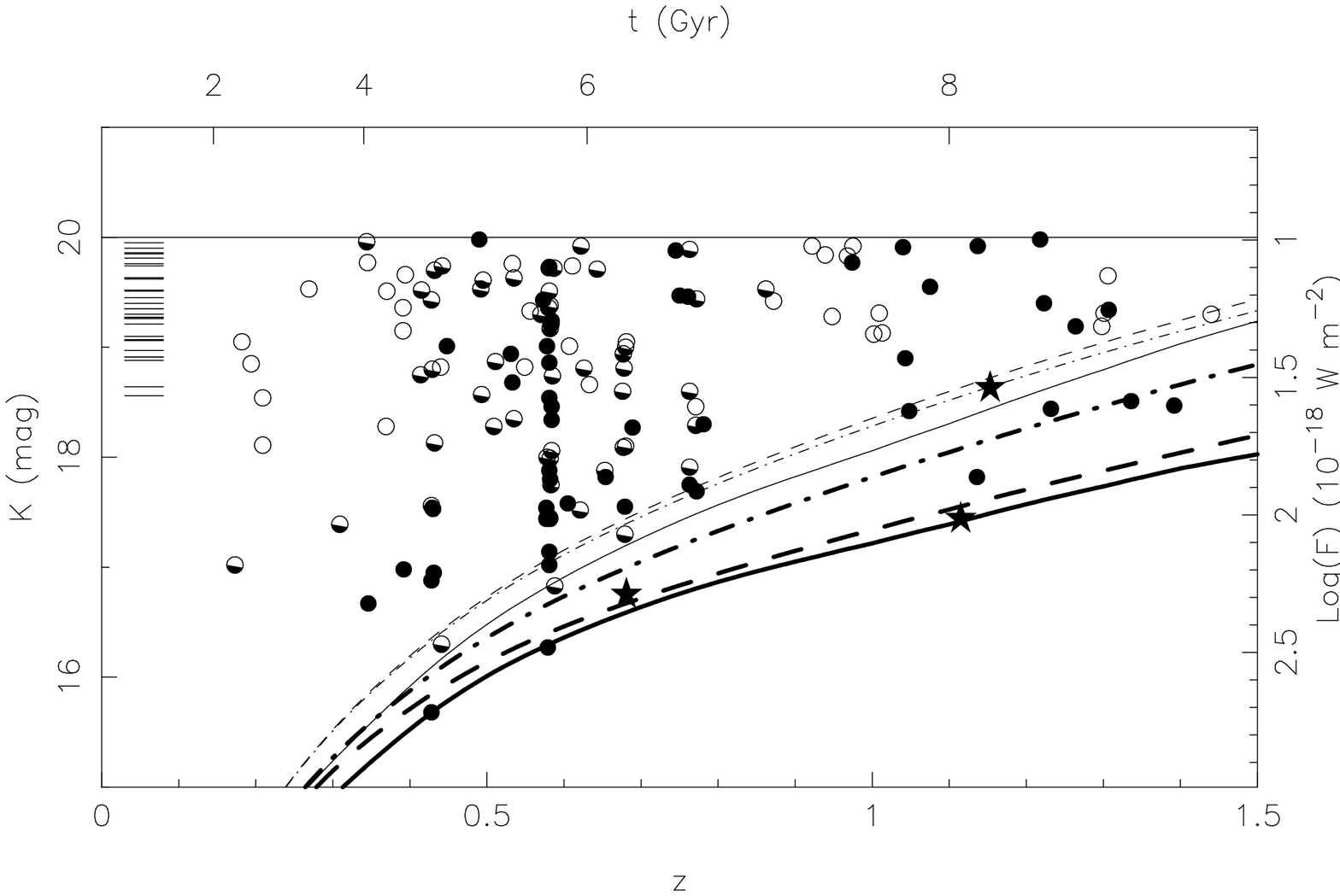}
\caption[figure1a.ps]{Hubble diagrams for the main sample 
for three observed magnitudes, 
Fig.~(a) $B$, Fig.~(b) $R$, Fig.~(c) $K$, using the
photometry of Pahre \etal\ (1998). The abscissa is the redshift 
$z$ rather than the conventional $\log z$. Also shown on the upper scale 
is the lookback time for our adopted cosmography.
The ordinate is the magnitude expressed as an equivalent flux in 
$\nu F_\nu$, measured in W m$^{-2}$ on the right hand scale. 
The open circles are ``${\cal E}$'' emission line 
galaxies, while the filled circles are ``${\cal A}$'' absorption line 
galaxies. ``${\cal C}$''
composite galaxies are desingated by half-filled circles and the 
three AGN as stars.
The measured magnitudes of the 32 remaining members of the galaxy 
sample for which spectroscopic redshifts are not available are shown as 
horizontal bars to the left of the diagrams. Note the firm upper 
limit on $K$ and the large variation
in color apparent in the broader distributions in $R$ and $B$.

Each of these diagrams includes theoretical evolutionary tracks
computed from the simulations of Poggianti (1997) 
(see text) under the assumption that an Sc Hubble type is equivalent to
an emission line galaxy (dot-dashed line), type Sa to a composite galaxy
(dashed line), and
type E to an absorption line galaxy (solid curve). 
The k-corrections alone produce
tracks for 2$L^*$ indicated by the lighter lines, while the thick lines
show the tracks for passive evolution. 
It is apparent that only at most a minority of galaxies, drawn from the
population, follow Poggianti's prediction for passive evolution. 
\label{fig1a}}
\end{figure}

\begin{figure}
\epsscale{0.35}
\plottwo{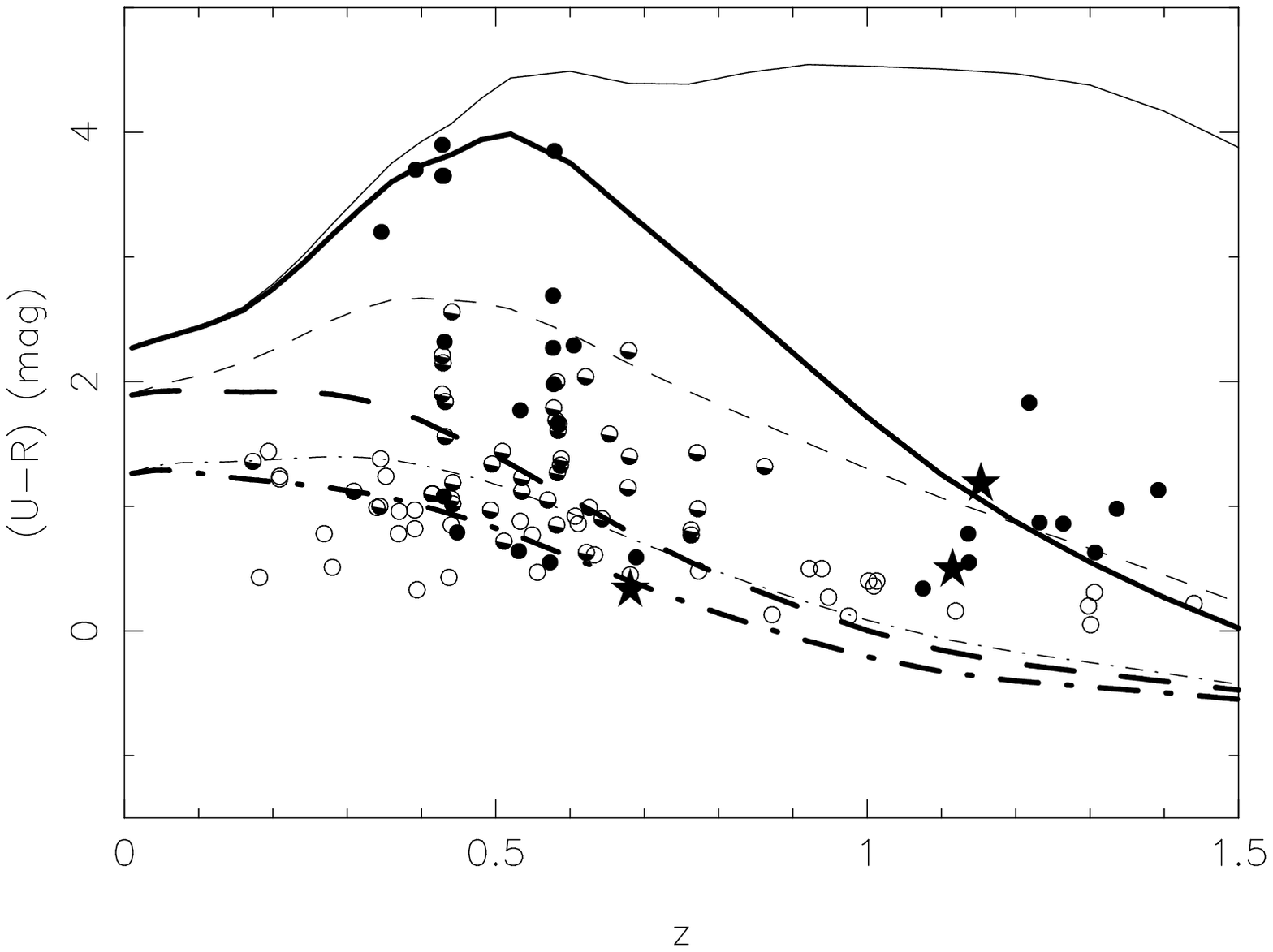}{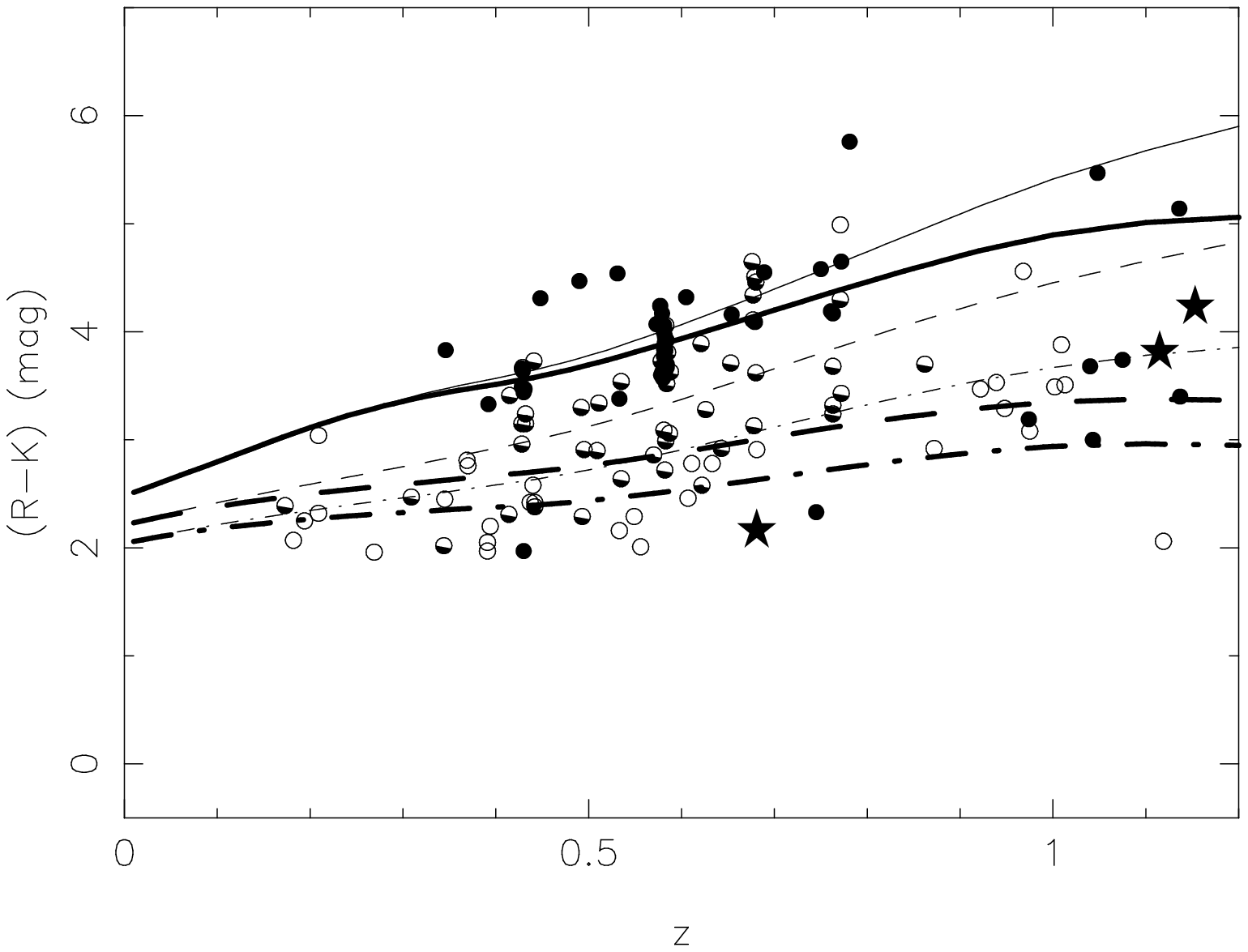}
\epsscale{0.70}
\plotone{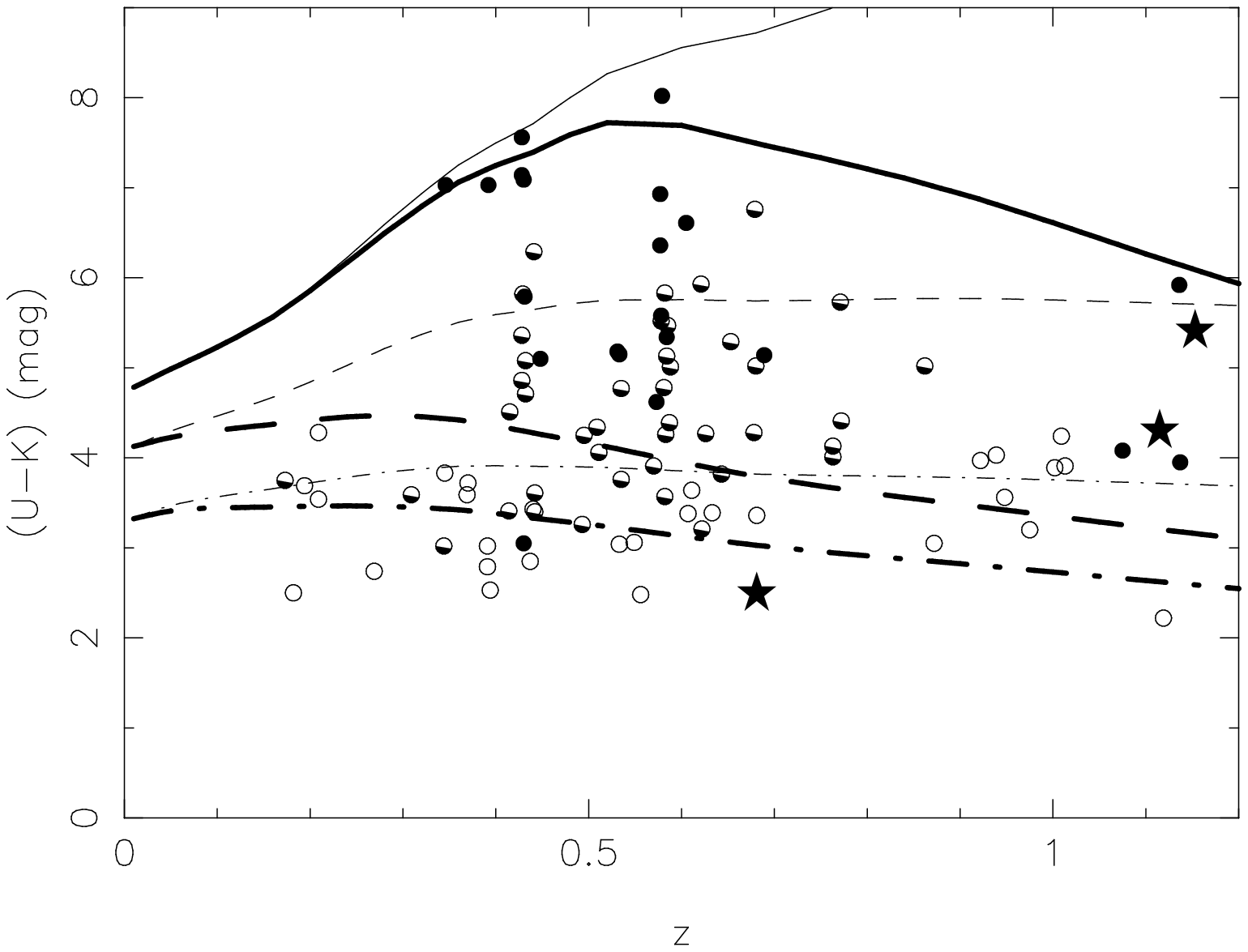}
\caption[figure2.ps]{The dereddened observed colors for the
extragalactic objects are shown as a function of redshift. 
The symbols indicating the various galaxy spectral types are the same
as in Figure~1.  The lines denote the predictions from the Poggianti models
for no-evolution and for passive evolution for
elliptical, Sa, and Sc galaxies as in Figure~1.  
The three panels show
$(U-R)$ (figure 2a), $(R-K)$ (figure 2b), and $(U-K)$ (figure 2c).
Galaxies with upper limits for $U$ are not plotted in figures 2a,2c.

\label{fig2}}
\end{figure}

\begin{figure}
\epsscale{1.0}
\plotone{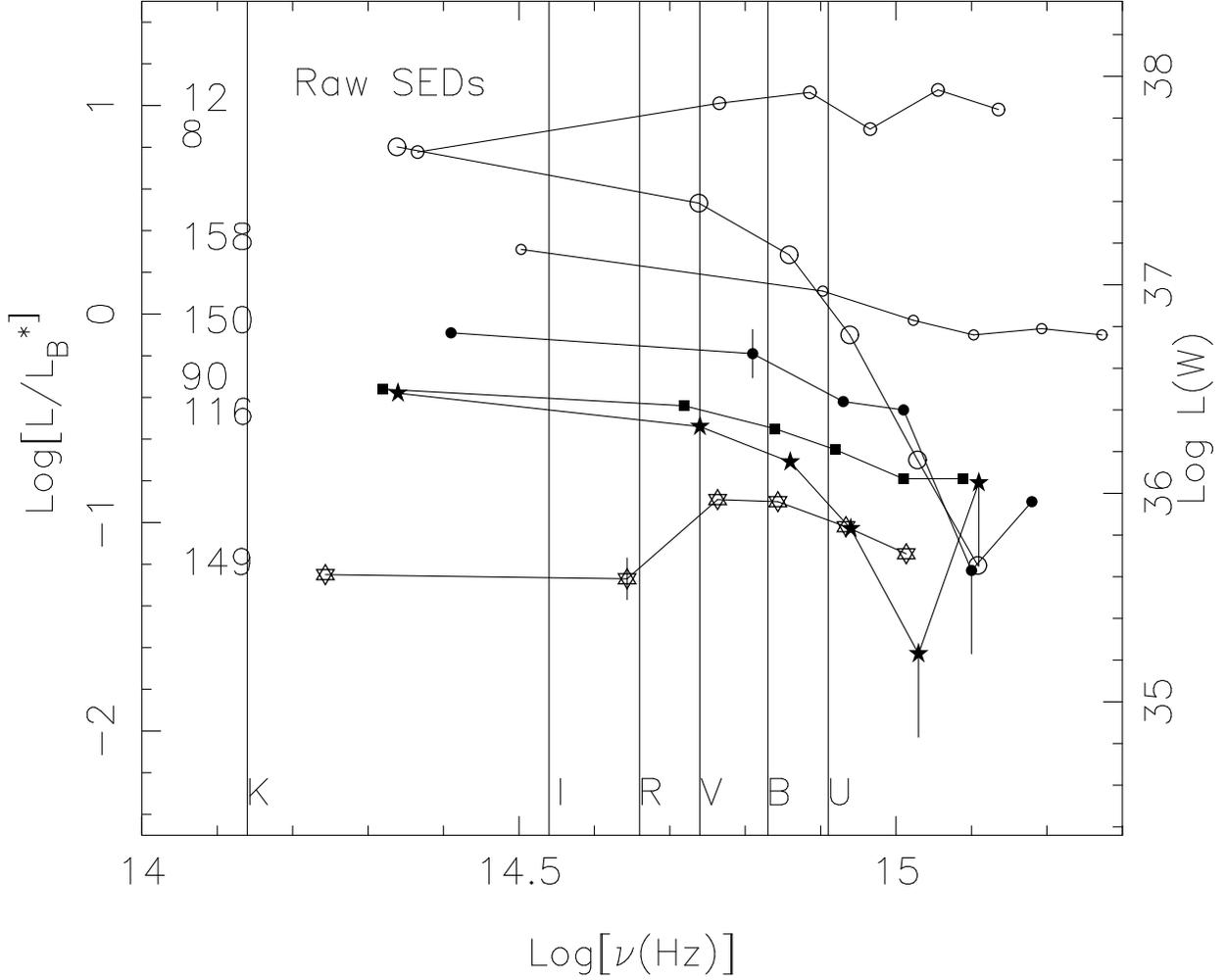}
\caption[figure3.ps]{Raw spectral energy distributions for a few 
selected galaxies
denoted by their identifying D0K numbers. The abscissa is the 
rest frequency in Hz. The ordinate is the spectral power
in units of the fiducial power $L_B^\ast$, (measured in W), where $\log(L_B^\ast)=36.86$.  
The errors in the measured magnitudes
are taken to be $0.2$ except where indicated by 
vertical lines extending above and below
the measurement.  Upper limits ($2\sigma$) are indicated by 
lines that extend downwards from the limit. 
In order to generate corrected SEDs, low accuracy measurements,
like the $I$ band measurement in D0K149 are replaced by interpolations.
Upper limits,
such as the $B$ and $U$ measurements in D0K116, 
are replaced by extrapolations through the first upper limit 
adopting it as a measured value. Galaxy D0K12 is an AGN and appears to vary.
\label{fig3}}
\end{figure}

\begin{figure}
\epsscale{0.4}
\plottwo{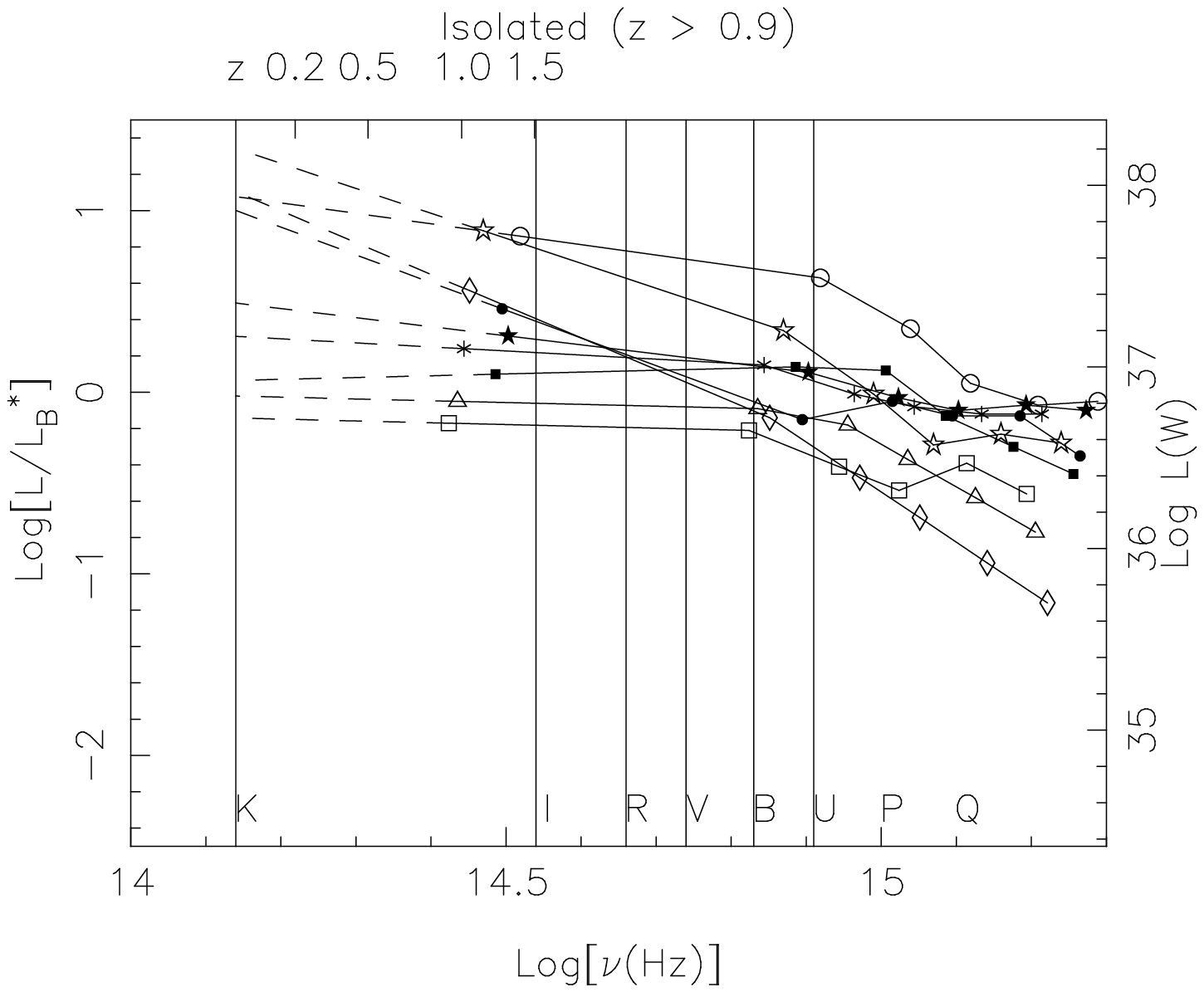}{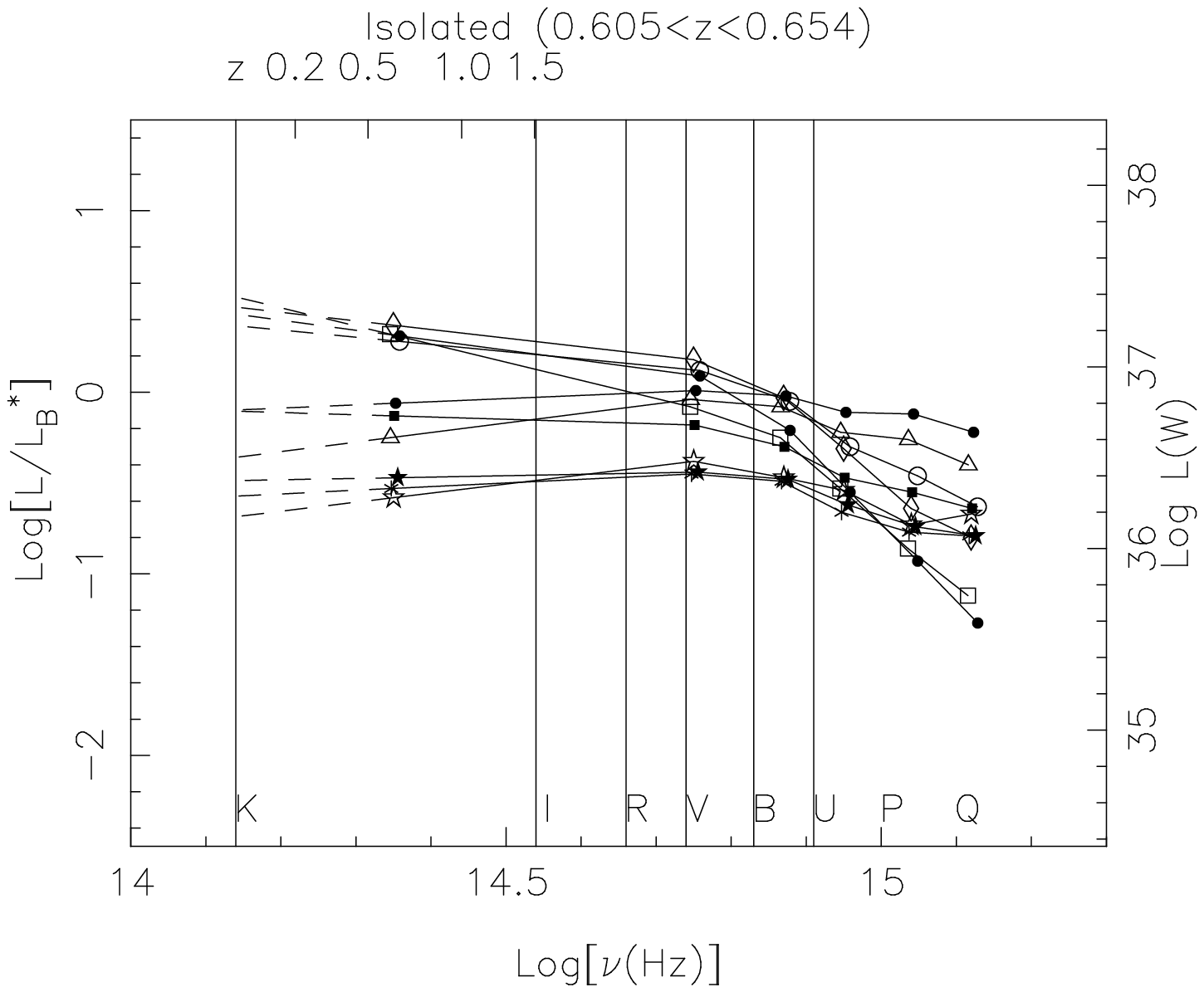}
\epsscale{0.4}
\plottwo{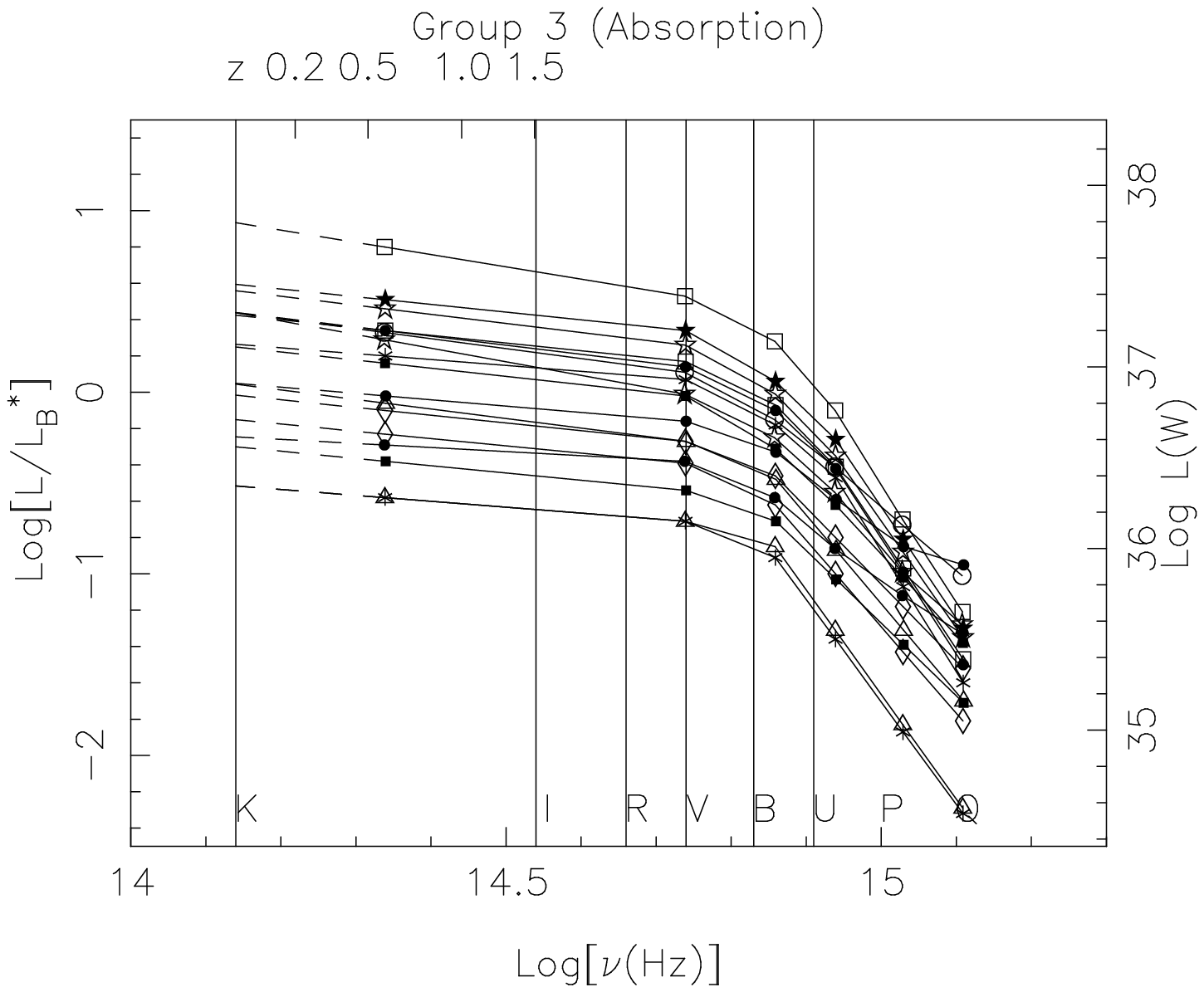}{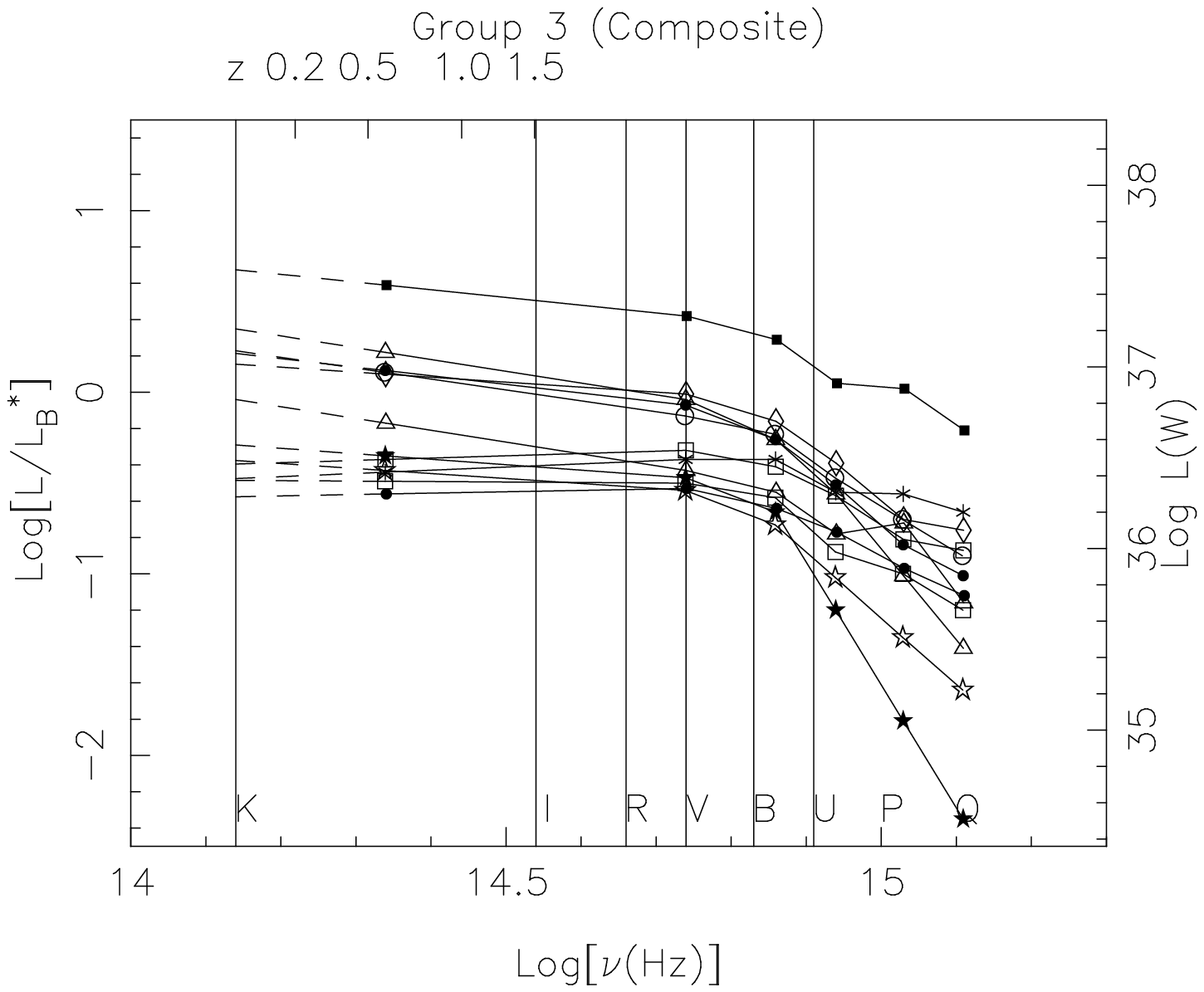}
\caption[figure4.eps]{Corrected rest frame SEDs for representative galaxies 
in the redshift sample.   The abscissa is the rest frequency and 
the rest wavelengths corresponding to our 6 color 
photometry augmented by the two supplementary ultraviolet bands 
$P$ and $Q$ are indicated.  The ordinate
is the logarithm of the spectral power in units of both 
$L_B^\ast$ and W.  Each galaxy
SED shows the rest wavelengths corresponding to the observations 
and dashed lines are used
to indicate extrapolations.  The upper horizontal scale can be used in conjunction with the $K$ point
to measure the redshift of the galaxy. 


The SEDs for selected galaxies (D0K183,172,108,64,42,188,113,158 and 68) 
with $z > 0.9$ shown in Figure~4a are remarkably blue. 
Figure~4b shows the SEDs for all galaxies in the region
$0.60 \le z \le 0.66$, a region outside the main redshift peaks.
By contrast, the SEDs for the absorption
line galaxies in the $z = 0.58$ peak shown in Figure~4c 
have quite red spectra.  The composite galaxies in this
redshift peak (Figure~4d) have composite
spectral properties.
(Spectral type ``${\cal EC}$'' 
galaxies are include with spectral type ``${\cal E}$''
and spectral type ``${\cal CA}$'' galaxies are included with
type ``${\cal A}$''.)
\label{fig4}}
\end{figure}

\begin{figure}
\epsscale{1.0}
\plotone{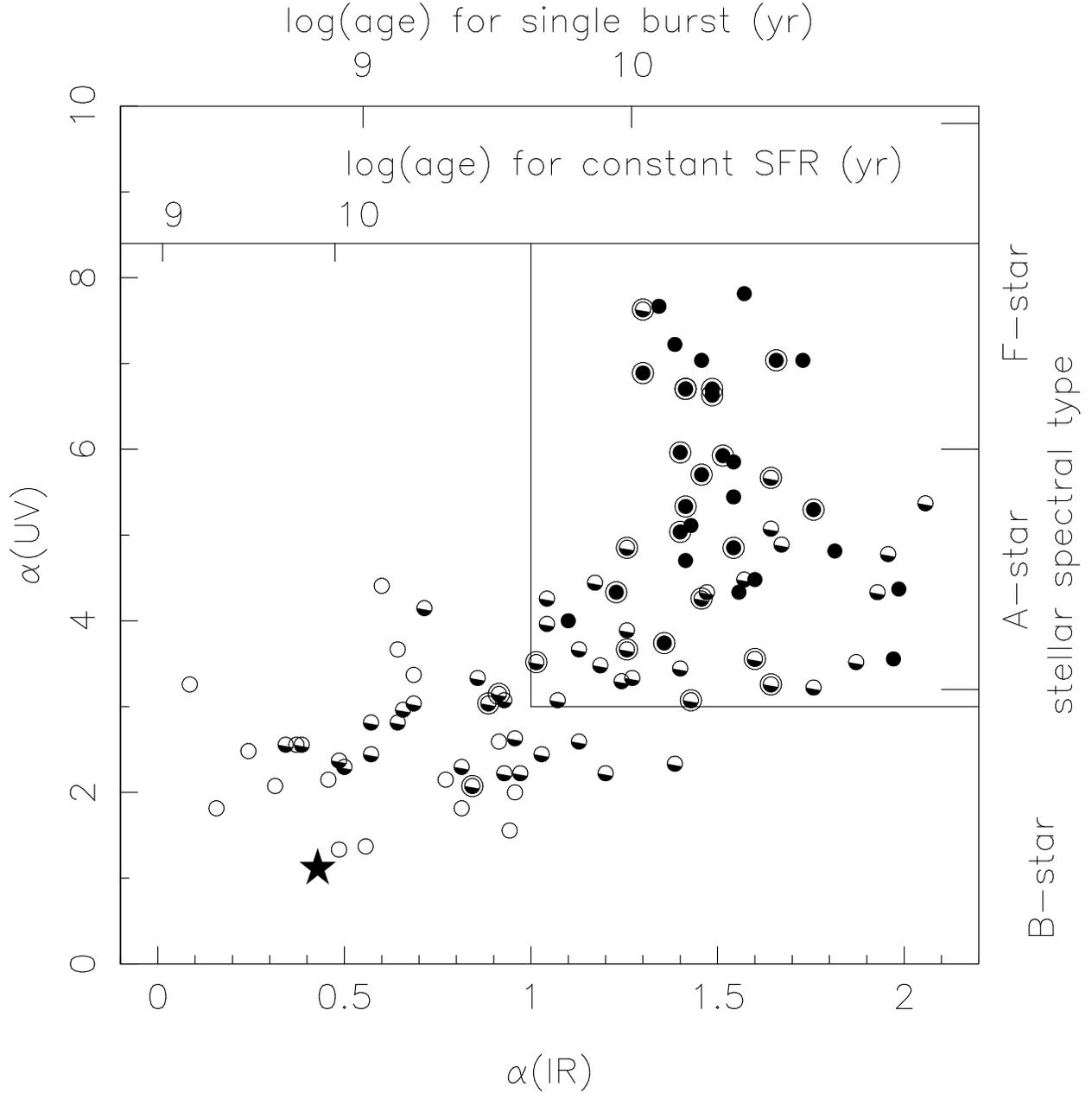}
\caption[figure5.ps]{Two color diagram that shows the clean spectral 
separation of the 
emission line galaxies (open circles) from the absorption line 
galaxies (filled circles).  The symbols for the galaxies belonging 
to Group 3 are circled.
Only galaxies with high quality redshifts
and $z < 0.8$ are displayed.
Composite galaxies (shown as half-filled circles) permeate both regions. 
The abscissa is the infrared color
expressed as a spectral index $\alpha_{IR}$. The ordinate
is an ultraviolet spectral index defined between the $B$ and 
$Q$  ($\log\nu=15.1$) bands.  The top axis gives the inferred age 
under two assumptions regarding the SFR, while the right axis
shows the spectral type characteristic of the UV light.
\label{fig5}}
\end{figure}

\begin{figure}
\epsscale{1.0}
\plotone{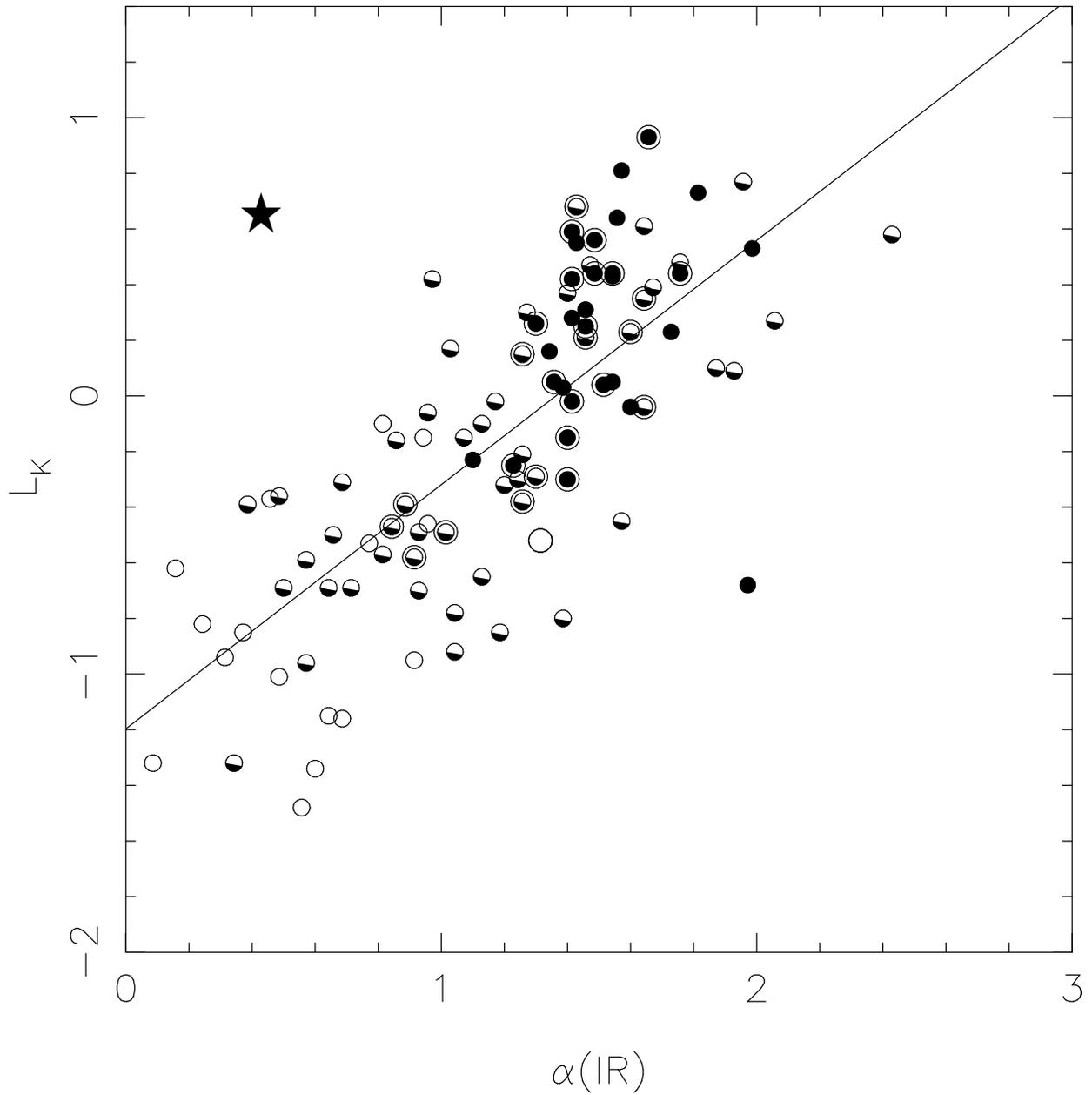}
\caption[figure6.ps]{The luminosity at $K$ 
is shown as a function of the infrared power law index $\alpha_{IR}$. 
The symbols and selection of galaxies displayed are the same as in Figure~5.
Note the separation of the galaxy spectral classes, emission
dominated galaxies being bluer and fainter.  The line represents
a linear least squares fit.
\label{fig6}}
\end{figure}

\begin{figure}
\epsscale{1.0}
\plotone{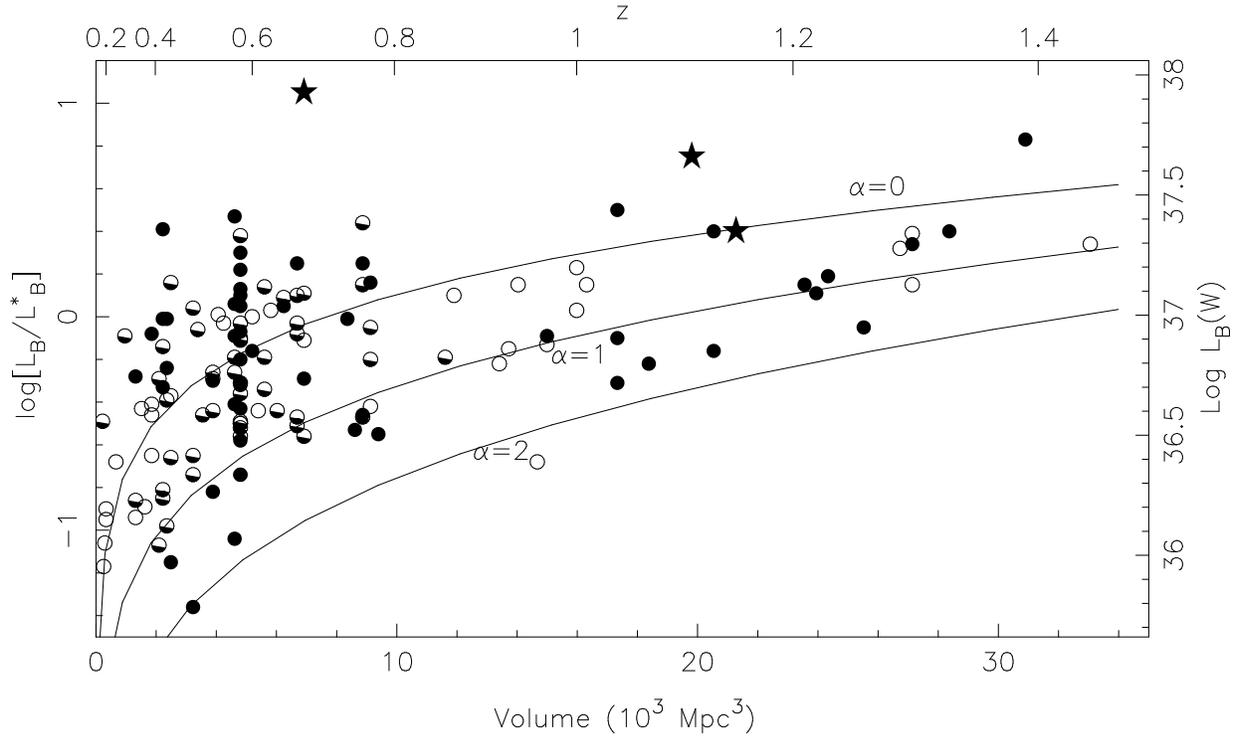}
\caption[figure7.ps] {The luminosity $L_B$ in units of 
$L_B^{\ast}$ is shown as a function of cosmological volume.
The corresponding redshift is shown at the top, while at
the right $L_B$ is given in units of W.  The symbols for the
galaxies are the same as in Figure~1.  The lines denote the
survey cutoff at an apparent magnitude of $K = 20$ 
for galaxies with $\alpha_{IR} = 0, 1, 2$.
\label{fig7}}
\end{figure}

\begin{figure}
\epsscale{1.0}
\plotone{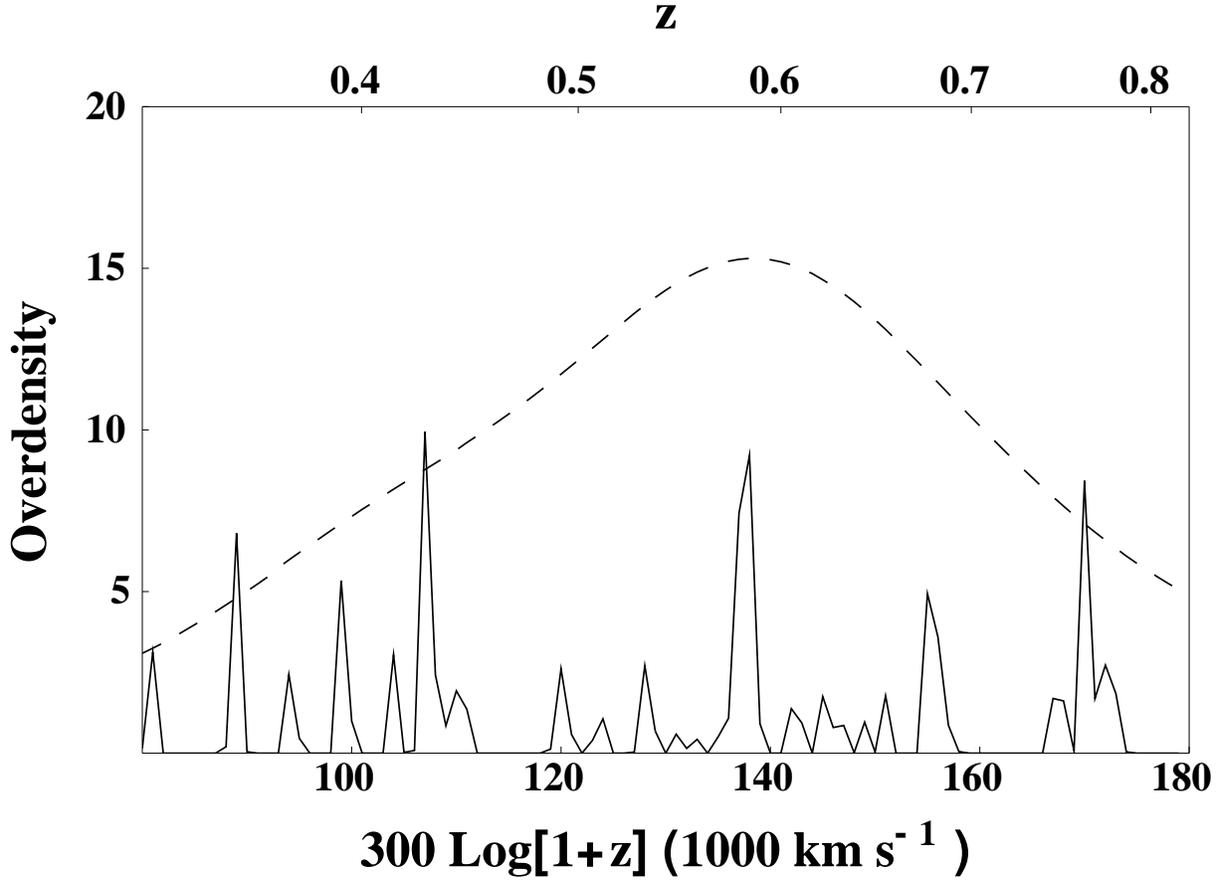}
\caption[figure8.ps]{Identification of galaxy groups in terms of overdensities.
The galaxy density in local velocity space is computed after smoothing the observed
distribution using a Gaussian with $\sigma=15,000$~km s$^{-1}$
(dashed curve).  The galaxy 
distribution is then recomputed using $\sigma=300$~km s$^{-1}$and the overdensity
relative to the smooth distribution is plotted (solid curve). The five groups
discussed in the text are clearly seen, along with evidence of strong clustering
in smaller groups.
\label{fig8}}
\end{figure}

\begin{figure}
\epsscale{1.0}
\plotone{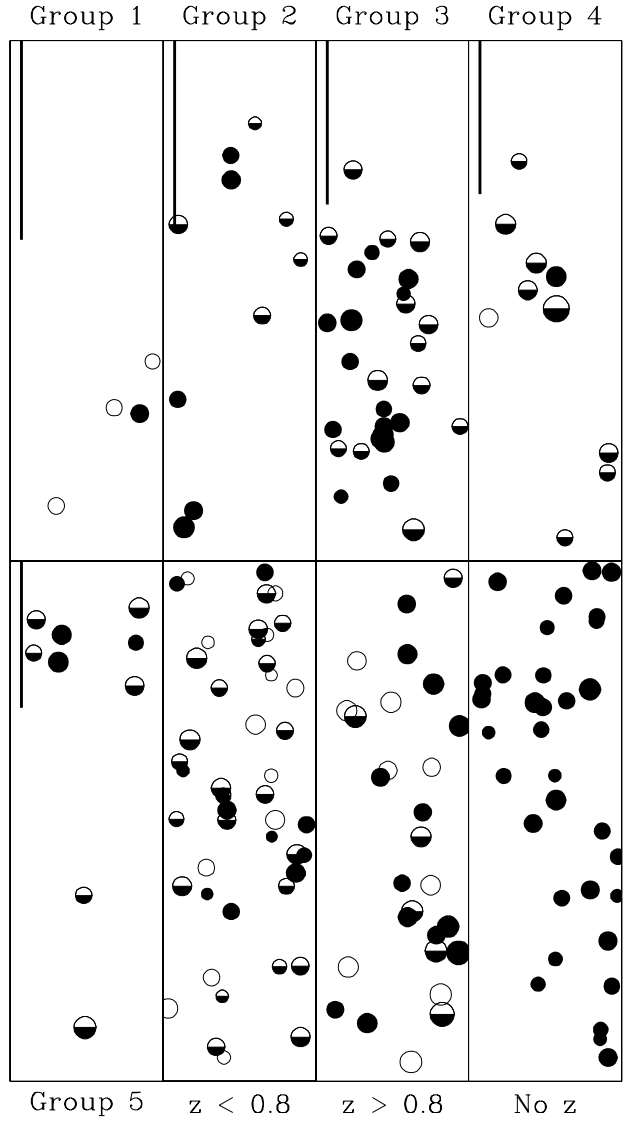}
\caption[figure9.eps]{The spatial distribution of each of the galaxies in
each of the five strongest redshift peaks is shown.  A 1 Mpc bar
is shown in each case. This is followed
by the spatial distribution of the low and high (cut at $z = 0.8$) $z$
objects not members of the five principal redshift peaks.  The final
panel shows the spatial distribution of the 34 objects without
redshifts.  The symbols used to denote galaxy type are the same
as in Figure~1.  The size of the symbols increases with the
apparent brightness.
\label{fig9}}
\end{figure}

\begin{figure}
\epsscale{1.0}
\plotone{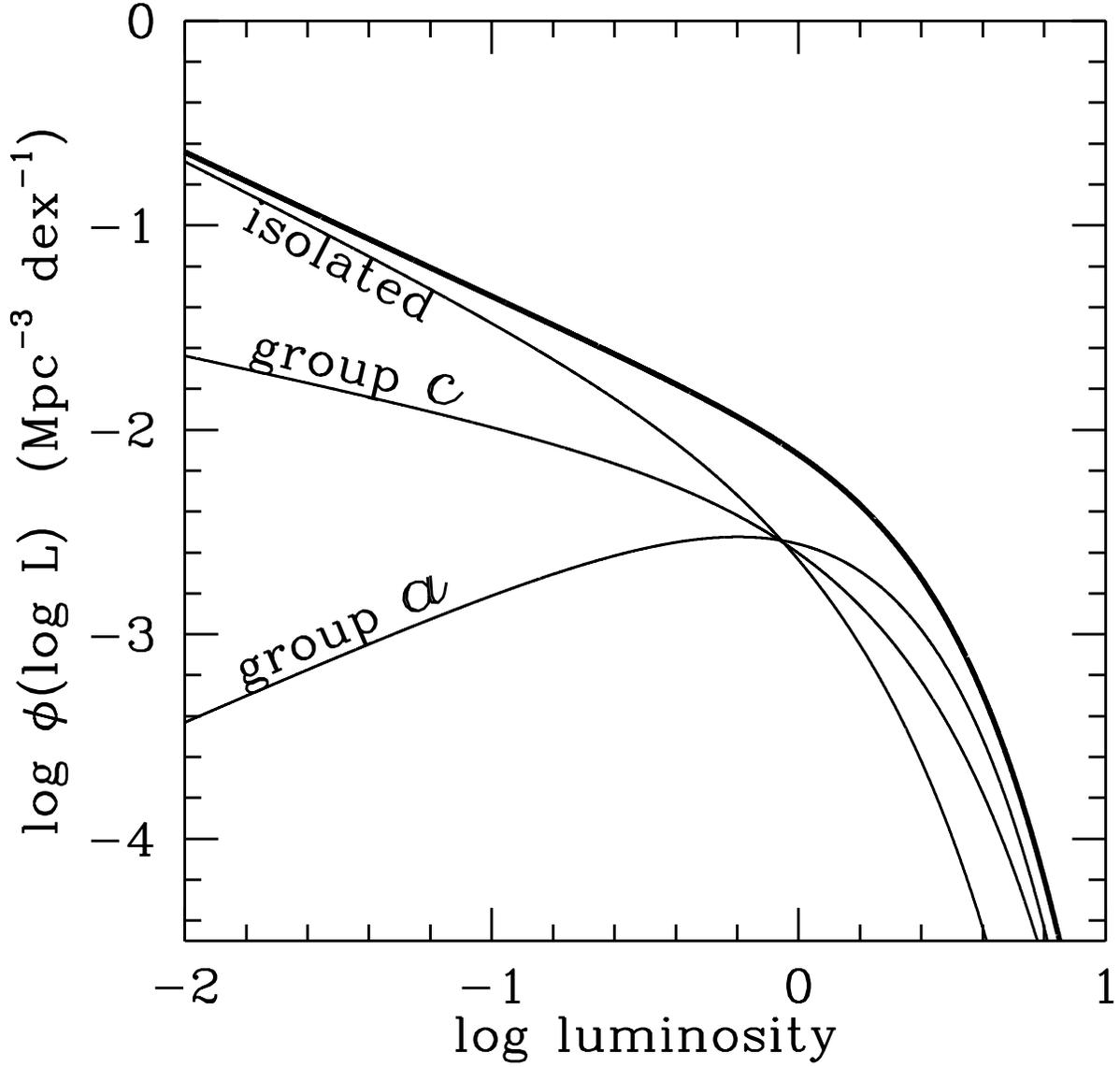}
\caption[figure10.ps]{The luminosity function inferred from the
galaxies in groups 2-5 of galaxy spectral type ``$\cal{A}$'' and of
galaxy spectral type ``$\cal{C}$'' (plus one galaxy of type ``$\cal{E}$'')
is shown.  The luminosity function for isolated galaxies with
$z < 0.8$ is also shown, as is the total luminosity function.
Note the differences in the faint end slope parameter $\alpha$ between
the various galaxy spectral groups.
\label{fig10}}
\end{figure}

\begin{figure}
\epsscale{1.0}
\plotone{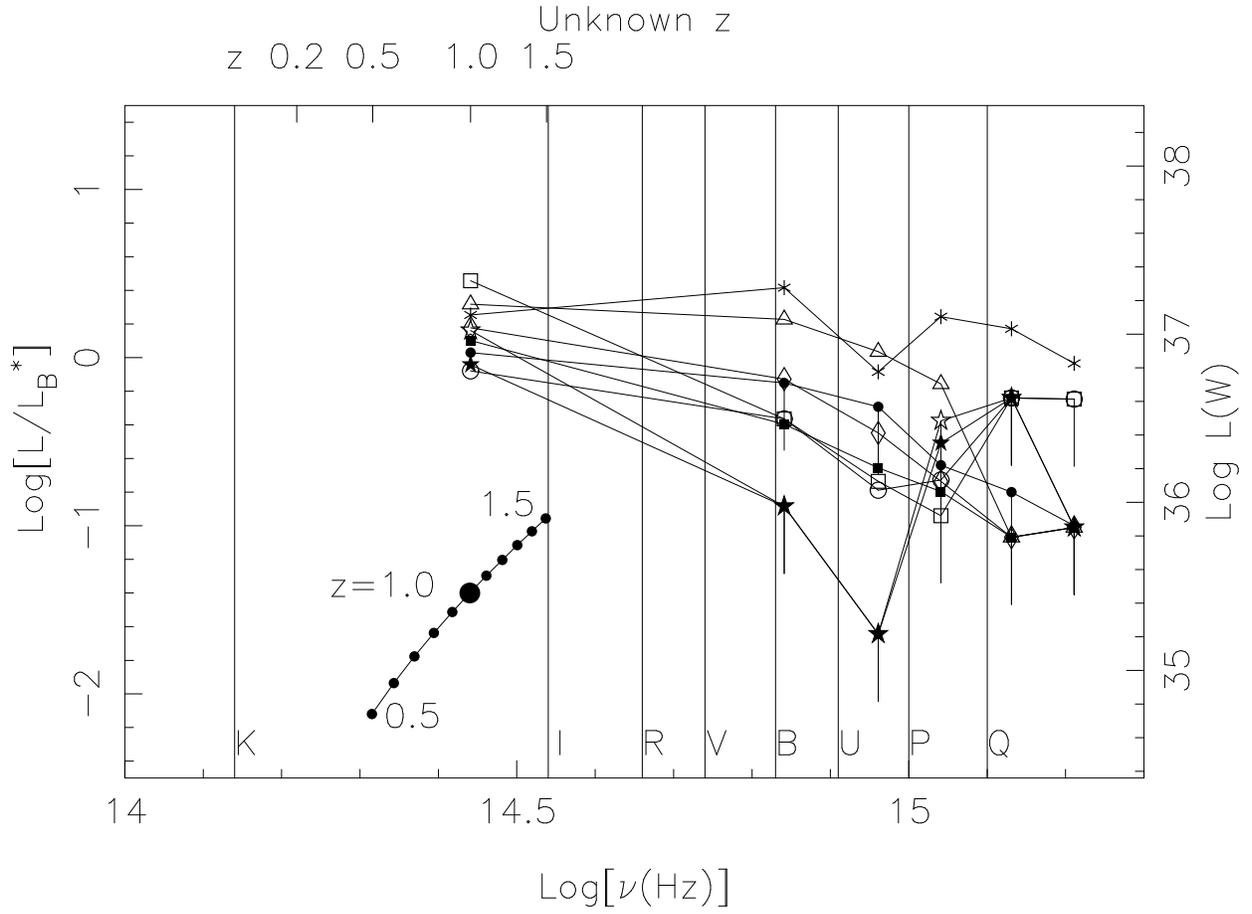}
\caption[figure11.ps]{The rest frame SEDs for a selection
of the 34 galaxies with no redshifts (D0K72, 494, 506, 117, 124, 141,
155, 173, and 180)
calculated assuming
$z = 1$ are shown.  The axes are the same as in Figure~4.  
The line in the lower left indicates how the SEDs will shift for 
$0.5<z<1.5$.
\label{fig11}}
\end{figure}

\end{document}